\documentclass[sigplan,10pt]{acmart}
\copyrightyear{2026}
\acmYear{2026}
\setcopyright{cc}
\setcctype{by}
\acmConference[EUROSYS '26]{21st European Conference on Computer Systems}{April 27--30, 2026}{Edinburgh, Scotland Uk}
\acmBooktitle{21st European Conference on Computer Systems (EUROSYS '26), April 27--30, 2026, Edinburgh, Scotland Uk}
\acmDOI{10.1145/3767295.3803612}
\acmISBN{979-8-4007-2212-7/2026/04}

\newcommand{\Paragraph}[1]{\noindent\textbf{#1}}
\newcommand{\Item}[0]{\noindent $\bullet$}

\usepackage{fontawesome5}

\usepackage{makecell}
\usepackage{colortbl}
\usepackage{multirow}

\usepackage{amsfonts}
\usepackage{xspace}
\usepackage{enumitem}

\newcommand{\sys}{\textsc{TAO}\xspace} 
\newcommand{\Hsh}{\mathsf{H}}           

\begin{document}
\title{TAO: Tolerance-Aware Optimistic Verification for Floating-Point Neural Networks}


\author{Jianzhu Yao}
\affiliation{%
  \institution{Princeton University}
  \city{Princeton}
  \state{NJ}
  \country{USA}}
\email{jy0246@princeton.edu}

\author{Hongxu Su}
\affiliation{%
  \institution{HKUST (GZ)}
  \city{Guangzhou}
  \country{China}}
\email{hsu238@connect.hkust-gz.edu.cn}

\author{Taobo Liao}
\affiliation{%
  \institution{University of Illinois Urbana-Champaign}
  \city{Urbana}
  \state{IL}
  \country{USA}}
\email{taobol2@illinois.edu}

\author{Zerui Cheng}
\affiliation{%
  \institution{Princeton University}
  \city{Princeton}
  \state{NJ}
  \country{USA}}
\email{zerui.cheng@princeton.edu}

\author{Huan Zhang}
\affiliation{%
  \institution{University of Illinois Urbana-Champaign}
  \city{Urbana}
  \state{IL}
  \country{USA}}
\email{huan@huan-zhang.com}

\author{Xuechao Wang}
\authornote{Corresponding authors.}
\affiliation{%
  \institution{HKUST (GZ)}
  \city{Guangzhou}
  \country{China}}
\email{xuechaowang@hkust-gz.edu.cn}

\author{Pramod Viswanath}
\authornotemark[1]
\affiliation{%
  \institution{Princeton University}
  \city{Princeton}
  \state{NJ}
  \country{USA}}
\email{pramodv@princeton.edu}

\renewcommand{\shortauthors}{Yao, Su, Liao, Cheng, Zhang, Wang, and Viswanath}

\begin{abstract}
Neural networks increasingly run on hardware outside the user’s control (cloud GPUs, inference marketplaces, edge specialized accelerators) for both training and inference. Yet ML-as-a-Service reveals little about \emph{what} actually ran or \emph{whether} returned outputs faithfully reflect the intended inputs and models. Users lack recourse against service downgrades such as model swaps, quantization, graph rewrites, or discrepancies like altered advertisement embeddings. Verifying outputs is especially difficult because floating-point execution on heterogeneous accelerators is inherently nondeterministic. Existing approaches like zkML, deterministic replay, TEEs, and replication are either impractical for real floating-point neural networks or reintroduce the need to trust the vendor. We present \sys: a \textbf{T}olerance-\textbf{A}ware \textbf{O}ptimistic verification protocol for floating-point neural networks that accepts outputs within principled \emph{operator-level acceptance regions} rather than requiring bitwise equality. \sys combines two complementary error models: (i) sound per-operator IEEE-754 worst-case bounds and (ii) tight empirical percentile profiles calibrated across hardware types. Discrepancies are resolved via a Merkle-anchored, threshold-guided interactive dispute game that recursively partitions the traced computation graph until one operator remains; at the leaf, adjudication reduces to either a lightweight theoretical-bound check or a small honest-majority vote against empirical thresholds. Unchallenged results finalize after a challenge window, without requiring trusted hardware or deterministic kernels.

We implement \sys as a PyTorch-compatible runtime and a coordinator contract layer; in our prototype, the coordinator is instantiated as an Ethereum smart-contract deployment (Holesky testnet) to quantify coordination cost. The runtime instruments graphs, computes bounds for each operator in real time, and runs unmodified vendor kernels in FP32 with negligible overhead (0.3\% additional latency on Qwen3-8B). Our evaluation spans CNNs, Transformers, LLMs, and diffusion models across multiple GPUs (A100, H100, RTX 6000, RTX 4090). Results show that empirical thresholds are $10^2$–$10^3\times$ tighter than theoretical bounds, and bound-aware adversarial attacks achieve \textbf{0\%} success under empirical checks. Together, \sys reconciles scalability with verifiability for real-world heterogeneous ML compute.
\end{abstract}



\maketitle

\section{Introduction}
\label{sec:introduction}

Modern machine learning (ML) systems increasingly outsource training and inference to third-party infrastructure, including public clouds, inference marketplaces, edge clusters, and specialized accelerators~\cite{zhang2019empirical}. This outsourcing is driven by the rapidly growing cost of ML computation, which has led to two notable trends. First, to conserve scarce compute budgets, practitioners often trade accuracy for cost at serving time, for example, by switching to smaller models, lowering precision, or coarsening quality knobs. Second, \emph{platform opacity} makes these trade-offs invisible: in ML-as-a-service (MLaaS) platforms, whether closed-source proprietary models or open-model inference services, users cannot see which model was executed, what inputs were processed, or whether results faithfully reflect the intended computation. As a result, users accept outputs they cannot independently verify, leaving them exposed to silent downgrades such as model swaps, early exit~\cite{lahiany2022pteenet}, or quantization~\cite{zhao2024atom}, as well as service-level discrepancies such as manipulated advertisement embeddings that redirect ad placements toward irrelevant products. When consumers of results do not control the underlying hardware, \emph{verifiable ML} becomes a crucial system problem: the user should be provided with an efficiently verifiable proof certifying that the output was generated by executing an agreed model on agreed inputs~\cite{jia2021proof, choi2023tools}.

\begin{figure}
\centering
\includegraphics[width=\linewidth]{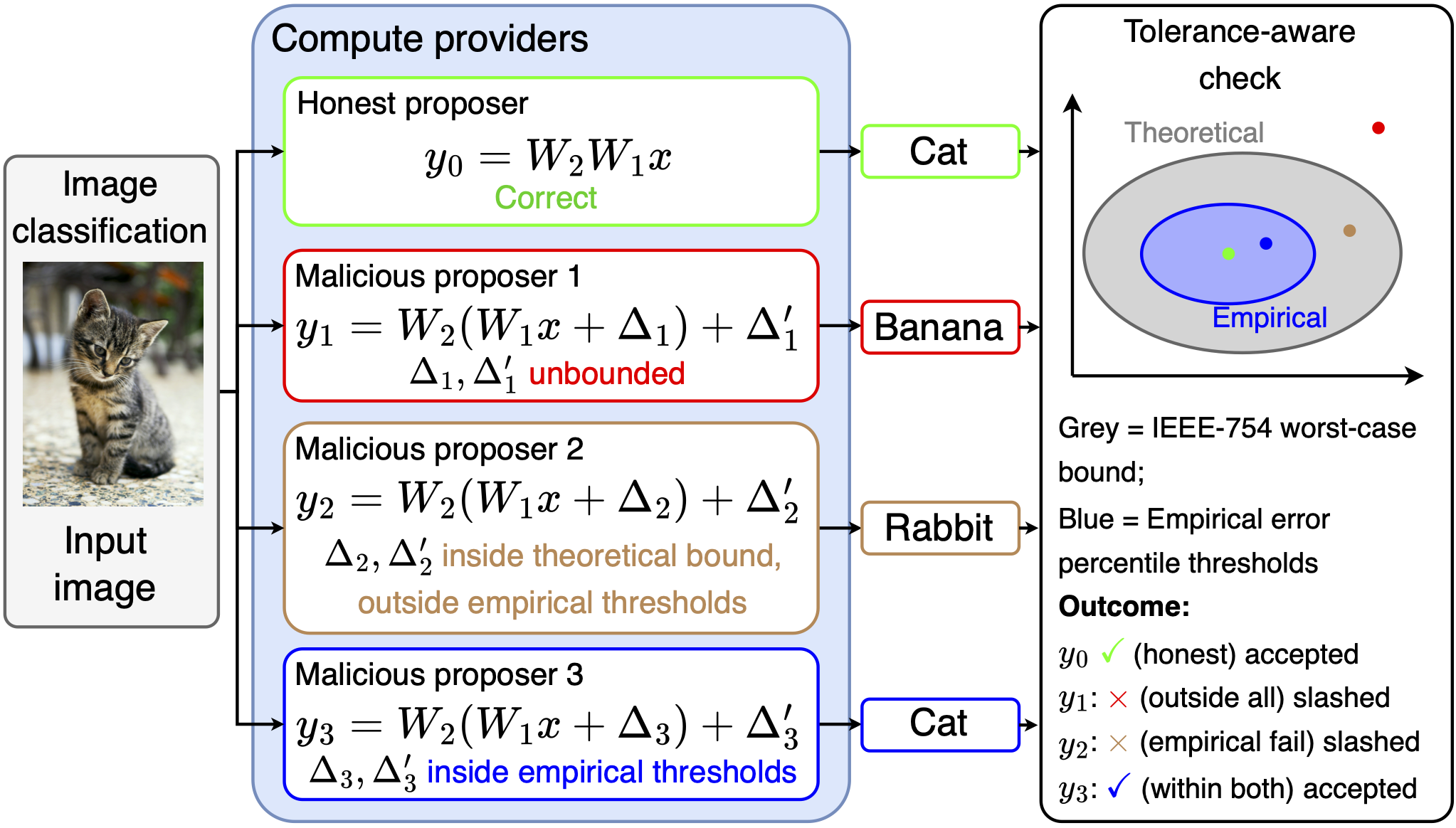}
\vspace{-20pt}
\caption{Tolerance-aware optimistic verification for floating-point NN inference: the honest proposer outputs ``Cat''; malicious proposers inject perturbations after each operator. Results are rejected if they fall outside IEEE-754 theoretical bounds or empirical thresholds. Bounded deviations that do not alter the final label are accepted.}
\label{fig:intro}
\Description{An illustration of tolerance-aware optimistic verification for image classification. The honest proposer outputs Cat; malicious proposers inject perturbations that are checked against IEEE-754 theoretical bounds and empirical thresholds.}
\vspace{-20pt}
\end{figure}

As demand for ML compute grows, platforms with decentralized inference markets and heterogeneous GPU fleets promise greater capacity and price diversity. However, they also expose a fundamental technical obstacle for verifiable ML: modern accelerators do not guarantee deterministic outputs. Floating-point arithmetic (IEEE-754)~\cite{kahan1996ieee} is non-associative; vendor kernels legitimately reorder reductions and fuse operators; thread scheduling and atomics introduce nondeterministic execution traces~\cite{chou2020deterministic,shanmugavelu2024impacts}. 
Consequently, running the same neural network (NN) on different hardware, or even repeated executions on the same GPU can yield slightly different outputs. Here, we use \emph{operator} to denote a primitive tensor-level function (a node in the NN graph), such as matrix multiplication, convolution, element-wise addition, or activation in PyTorch/ATen. In short, \emph{cross-platform nondeterminism is intrinsic to production GPU stacks}, yet most existing verification methods assume exact reproducibility. This work addresses the gap.

\Paragraph{Why existing approaches fall short.}
(1) \emph{zkML}: A natural path is to prove every arithmetic operation in \emph{zero knowledge}. However, current end-to-end zkML pipelines remain orders of magnitude slower and memory-intensive for realistic floating point models \cite{wang2023ezdps,chen2024zkml,qu2025zkgpt,sun2024zkllm}. Expressing large PyTorch graphs in field arithmetic or fixed-point circuits, and proving native FP32 GPU kernels at scale, is beyond today's practicality. Moreover, real GPU stacks are inherently nondeterministic: one must either fix a specific execution schedule (sacrificing performance and hardware heterogeneity) or encode acceptance intervals, neither of which current zkML frameworks support. (2) \emph{Deterministic replay}: Another approach is to enforce deterministic libraries, kernels, and reduction orders so that independent re-executions are bitwise reproducible~\cite{zheng2021agatha}. This discards mature vendor libraries, requires extensive manual redesign of parallel reductions, and constrains accelerators (GPUs/TPUs) that are fundamentally nondeterministic by design. (3) \emph{Trusted execution environments (TEEs)} attest that a computation ran inside a trusted enclave~\cite{tramerslalom,moon2025asgard,bai2025phantom}. But this shifts trust to single-vendor microcode, introduces side-channel risks~\cite{wei2018know} and performance degradation~\cite{niu20223legrace} due to limited storage and compute capacity. (4) \emph{Replication and majority voting}: Protocol-level schemes replicate the computation across multiple providers and accept the majority output. The high cost of replication for large models severely limits practicality and scalability~\cite{zhao2005result,esmaeili2024serene}. In summary, none of these approaches simultaneously accommodate floating-point nondeterminism, preserve production-level performance, and enable portable, vendor-agnostic verification.

\Paragraph{Our approach: verify up to principled error thresholds.}
NN models are inherently robust to small rounding differences, and user-visible semantics rarely depend on bit-exact equality. Thus, the appropriate notion of correctness is \emph{approximate}: an output is acceptable if it falls within a model- and operator-specific tolerance induced by IEEE-754 arithmetic. The key challenge is to formalize this tolerance precisely, make it compositional across operators, and ensure that it can be verified efficiently.

\Paragraph{This paper: \sys.}
We present \sys, a \emph{tolerance-aware optimistic execution} protocol for verifiable neural network computation on heterogeneous accelerators. Instead of requiring bitwise equality, \sys verifies that operator-level outputs fall within prescribed, model-specific error bounds. Discrepancies are resolved through a Merkle-anchored interactive dispute game that localizes disagreements to a single operator (a leaf operator in the Merkle tree~\cite{merkle1987digital}). At that leaf, lightweight adjudication suffices, either through a certified theoretical bound check or a small committee vote against pre-calibrated empirical thresholds. The key idea is to embrace, rather than eliminate, floating-point nondeterminism by verifying neural outputs \emph{up to} principled per-operator error bounds, while preserving native GPU performance and hardware heterogeneity. By pushing verification down to the granularity where it is both meaningful and efficient: \sys reconciles scalability with accountability. Users can run the models they want, on any hardware the market provides, without compromising correctness, transparency, or recourse.

\Paragraph{Two complementary error models.}
Even when software determinism is enforced (e.g., fixed kernel choices, disabled autotuning), parallel execution still reorders operations, introducing nondeterminism due to the non-associativity of IEEE-754 arithmetic. Rather than attempting to eliminate this through brittle execution constraints, we model and bound rounding error directly. \sys introduces two complementary error models that make tolerance-aware verification both principled and deployable:

\Item{} \emph{Theoretical IEEE-754 bounds.} For each operator, we compute a worst-case element-wise rounding error as a function of its inputs. This bound is sound by construction but overly conservative in deep networks, so we apply it only at the leaf level, where it remains both cheap and decisive.

\Item{} \emph{Empirical error percentile thresholds.} We calibrate rounding errors offline across accelerators to capture the \emph{distribution} of cross-hardware deviations for each operator. These empirical profiles are tight, model-specific, scalable, and only need to be calibrated once (easy to post-verify). We use them to guide the search for the disputed leaf and, when necessary, to adjudicate with a small honest-majority committee, where they are more expensive than theoretical bounds but also significantly tighter and more robust.

Our traced runtime instruments PyTorch graphs~\cite{ansel2024pytorch}, computes theoretical bounds on the fly, and records intermediate results to build robust cross-device empirical percentile profiles. Instead of propagating worst-case error across the entire graph, we \emph{turn composition into localization}: disagreements are recursively narrowed until a single operator can be certified. By combining both error bounds, \sys flexibly balances universality, tightness, and scalability.

\Paragraph{From whole-model proofs to single-operator checks.}
Verifying a large model end-to-end against any bound remains expensive. \sys makes verification \emph{cheap} by localizing disagreement and reducing the verification burden. Model owners provide cryptographic commitments to the weights and graph topology. In the common case, a proposer (compute provider) runs the model and posts a commitment; if unchallenged within a window, the commitment attains \emph{finality}. If challenged, an interactive $N$-way partition protocol is invoked. At each round, empirical thresholds guide the split: the proposer partitions the traced graph into disjoint subgraphs with inclusion proofs, and the challenger identifies the first child whose outputs exceed its empirical thresholds. After $O(\log_{N}|V|)$ rounds ($V$ is the set of operators), the interactive dispute game narrows to a single operator with agreed inputs, where a theoretical-bound check or a committee vote evaluation resolves the dispute.

\Paragraph{Coordination layer.} The protocol needs a coordination service that (a) stores tamper-evident commitments to model, graph, and outputs; (b) manages challenge windows and state transitions; and (c) escrows bonds and enforces rewards and punishments. One straightforward option is to deploy a trusted centralized coordinator. In our prototype, however, we instantiate this layer on a decentralized ledger (Ethereum), which provides authenticated logs, fair timeouts, and bond management out of the box, while also offering a transparent, widely understood cost model. Importantly, \sys itself does \emph{not} rely on blockchain-specific assumptions.

\Paragraph{Open-model setting and service transparency.}
\sys targets the prevalent and rapidly growing \emph{open-model} setting, where both weights and graph structure are public (e.g., community LLMs). This transparency enables subgraph extraction, per-operator thresholding, and provenance commitments. Closed-model APIs can also benefit, if providers expose such commitments to permissioned parties.

\Paragraph{Adversary and attack coverage.}
Tolerance-aware verification raises two concerns: \emph{trustworthiness} and \emph{attack surface} given the error bounds. We adopt a strong threat model in which the proposer may be malicious, injecting perturbations into intermediate tensors to evade empirical thresholds or theoretical error bound checks. To evaluate this risk, we design adaptive gradient-based attacks~\cite{madry2018towards}: distributional evasion against empirical envelopes and element-wise evasion against theoretical bounds. We further study whether many small, locally admissible deviations could accumulate into a task-level error without being detected. Our results show that the admissible per-operator perturbations are small and do not naturally compound, making end-to-end attacks under \sys practically difficult. An example is shown in Fig.~\ref{fig:intro}.

\Paragraph{Implementation overview.}
We implement \sys as a PyTorch runtime and a set of Ethereum smart contracts as the coordinator. The runtime instruments computation graphs, computes per-operator bounds on the fly, records intermediate traces, partitions and re-executes subgraphs, and emits or verifies Merkle commitments. Smart contracts manage commitments, dispute state transitions, per-round bonds, and misbehavior penalties. The runtime leverages software-level determinism features, runs unmodified vendor kernels in FP32, and uses FP64 arithmetic for error-bound calculations. It introduces negligible overhead in the optimistic execution ($\approx$0.3\% additional latency on \texttt{Qwen3-8B}) and requires no extra memory beyond native subgraph execution (Sec.~\ref{sec:implementation-and-evaluation}).

\Paragraph{Evaluation highlights.}
Across four representative models (\texttt{ResNet-152}, \texttt{BERT-large}, \texttt{Qwen3-8B}, \texttt{Stable Diffusion v1-5}) and four GPUs (RTX 4090, RTX 6000, A100, H100), we find empirical per-operator errors are $10^2\text{--}10^3\times$ tighter than worst-case IEEE~754 theoretical bounds for transformers (Fig.~\ref{fig:empirical-theoretical-comparison}). Under meticulously designed attacks, empirical error percentile thresholds yield a $0\%$ attack success rate (ASR) and $0\%$ false positives across models, even when thresholds are relaxed by a factor of $3$. By contrast, relying solely on theoretical bounds leaves a small but non-negligible gap: up to $2.4\%$ ASR on \texttt{Qwen3-8B}(Table~\ref{tab:main-threshold-scaling-full}). These results confirm that theoretical worst-case bounds can be overly loose, while empirical thresholds are both tight and robust, motivating committee-based adjudication at the leaf when needed. 

\Paragraph{Our contributions.}
\begin{enumerate}[leftmargin=*, topsep=0pt, itemsep=0pt, parsep=0pt]
  \item \textbf{Semantics for economically verifiable floating-point ML.} We formalize \emph{tolerance-aware} correctness for tensor programs and design an \emph{optimistic} challenge-response protocol that achieves economic finality without requiring determinism.
  \item \textbf{Error analysis and adversarial attacks at operator granularity.} We develop dual error models: portable theoretical IEEE-754 bounds and tight empirical percentile profiles, and conduct a comprehensive study of bound-aware adversarial attacks, quantifying their limited ability to subvert \sys across representative models.
  \item \textbf{Efficient and robust dispute resolution.} We design a Merkle-anchored, threshold-guided dispute game that recursively localizes disagreements to a single operator, supports verifiable subgraph provenance, and enables low-cost adjudication that respects nondeterminism via bound checks or honest-majority voting.
  \item \textbf{Deployable PyTorch runtime and end-to-end system.} We implement a PyTorch-compatible runtime that instruments models to compute and apply these error models in real time, integrate it into the full \sys pipeline, and demonstrate deployability in the open-model setting.
\end{enumerate}

\noindent {\bf Outline of the paper}. We first present our system architecture and protocol overview in Sec.~\ref{sec:architecture}, followed by detailed methodology of two error models in Sec.~\ref{sec:error-bounds}, and the adversary model and attack experiments in Sec.~\ref{sec:adversary-model}. We introduce the optimistic dispute game in Sec.~\ref{sec:methodology} and evaluate our system's performance, security, and cost in Sec.~\ref{sec:implementation-and-evaluation}.
Discussions are in Sec.~\ref{sec:discussion}, with related work in Sec.~\ref{sec:related-work} and conclusion in Sec.~\ref{sec:conclusion}.
\section{System architecture and protocol overview}
\label{sec:architecture}

\begin{figure*}[htbp]
\centering            
\includegraphics[width=\linewidth]{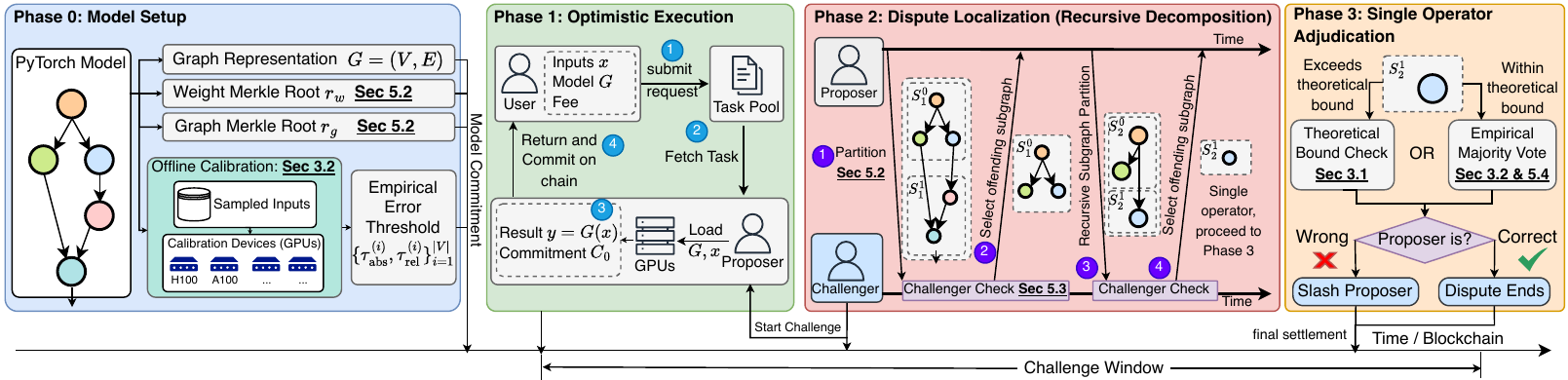}
\vspace{-24pt}
\caption{\sys overview. \textbf{Phase 0 (Model Setup)}: model owner commits to weights, graph, empirical error percentile thresholds.
\textbf{Phase 1 (Optimistic Execution)}: proposer posts a commitment for $(x,y)$.
\textbf{Phase 2 (Dispute Localization)}: an $N$-way, Merkle-anchored dispute game recursively narrows disagreement to a single operator following a canonical partition policy.
\textbf{Phase 3 (Single-Operator Adjudication)}: either a sound IEEE-754 error-bound check or a committee vote against empirical thresholds decides the outcome, triggering payment or punishment.}
\label{fig:system-architecture}
\Description{A diagram showing the workflow of the system. It consists of four phases: Model Setup, Optimistic Execution, Dispute Localization, and Single-Operator Adjudication. Each phase is described in detail, highlighting the key steps and interactions between the proposer, challenger, committee members, and users.}
\end{figure*}

In this section, we define the roles, assumptions, and the end-to-end protocol that makes NN results verifiable.

\Paragraph{System at a glance.}
\sys provides \emph{verifiable} results for NN workloads by combining optimistic dispute-based verification with \emph{tolerance-aware verification}. Fig.~\ref{fig:system-architecture} shows the four phases: in \emph{Phase~0 (Setup)}, the model owner commits Merkle roots of the weights and graph, and publishes calibrated per-operator empirical error percentile thresholds. In \emph{Phase~1 (Optimistic execution)}, a proposer runs the model to fulfill the user's request and the result finalizes after a short challenge window. If challenged, it runs a Merkle-anchored, threshold-guided interactive \emph{Phase~2 (dispute game)}, that recursively narrows disagreement from the entire model graph to a single PyTorch operator. Finally, \emph{Phase~3 (Single-operator adjudication)} executes either (i) a \textbf{sound, cheap} but potentially permissive IEEE-754 theoretical-bound check, or (ii) a \textbf{tighter but costlier} committee vote against the empirical error percentile thresholds.

\subsection{Roles and assumptions}

\Paragraph{Roles.} The system involves five roles:

\Item{} User: specifies model, submits inputs and requests results.

\Item{} Proposer: executes the model on its choice of hardware accelerators and posts the results with a commitment. If the result is questioned, it participates in dispute games. 

\Item{} Challenger: verifies result by re-executing the program. If its reproduced output deviates by the empirical error percentile threshold, it initiates and advances the disputes.

\Item{} Coordinator: an authenticated coordination service that records commitments, enforces timeouts, and orchestrates the state machine.

\Item{} Committee: sampled participants that adjudicate the final single operator against empirical error percentile thresholds when the theoretical check is inconclusive.


\Paragraph{Assumptions.} We assume IEEE-754 compliance with round-to-nearest-even; operator libraries may reorder reductions but must respect IEEE semantics. When invoked, the committee has an honest majority, and no trusted hardware is required. Most specific optimizations are compatible with \sys, with details discussed in Sec.~\ref{sec:discussion-optimization-compatibility}. We assume challengers have access to comparable compute resources as proposers, while the committee can have substantially less compute. Challengers are not assumed to be honest. 

\subsection{End-to-end protocol lifecycle}
\label{sec:system-architecture-and-workflow}

In this section, we walk through the four phases: from commitment to adjudication, showing how disputes are localized and resolved in a high-level overview (Fig.~\ref{fig:system-architecture}).

\Paragraph{Phase 0: Model setup.} At deployment time, the model owner serializes the PyTorch model into an acyclic dataflow graph \(G=(V,E)\) with a canonical topological order, where each node \(v_i\in V\) denotes a tensor operator (e.g., \textsc{MatMul}, \textsc{Softmax}, \textsc{GELU}) and each edge \((u,v)\in E\) captures a data dependency. The model state is then committed by merkleizing the weight tensors to obtain root \(r_w\) and, separately, merkleizing the graph topology (operator signatures) to obtain root \(r_g\). The pair \((r_w,r_g)\) is recorded on the coordinator backend, enabling \emph{lightweight provenance proofs} for any subgraph that arises during later dispute phases (Sec.~\ref{sec:graph-representation-and-verifiable-subgraph-extraction}). We also perform an offline calibration (Sec.~\ref{sec:empirical-error-percentile-threshold-calibration}) to quantify cross-hardware floating-point variability per operator, yielding empirical error percentile thresholds, which are also committed. These commitments enable \emph{verifiable subgraph extraction}: any subgraph \(S\subseteq G\) carries membership proofs into \(r_g\) and \(r_w\). 

\Paragraph{Phase 1: Optimistic execution (happy path).} Given input $x$ for model $M$, the proposer computes $y = G(x)$ with native devices and posts a commitment $C_{0}$ to the coordinator backend at time $t_{0}$: $C_0 = \text{H}(r_w||r_g||H(x)||H(y)||\text{meta})$, where $\Hsh(\cdot)$ is SHA-256 and ``meta'' encodes device type, kernel versions, dtypes, and the challenge window $\Delta$. If no challenge by $t_{0} + \Delta$, the coordinator will release user's payment to the proposer and that computation request is finalized.

\Paragraph{Phase 2: Dispute localization (interactive decomposition).} If any verifier recomputes $y' = G(x)$ and finds that the discrepancy between $y$ and $y'$ exceeds the empirical error percentile threshold, they may initiate a dispute game and act as a \emph{challenger}. The coordinator then freezes proposer's collateral $\sigma_{P}$, starts a dispute game, and initializes round $k{=}0$ with $S_0{:=}G$. Each round proceeds (details in Sec.~\ref{sec:iterative-subgraph-splitting}):

\Item{} \textbf{Partition (deterministic).} The proposer partitions the current disputed subgraph $S_k$ into $N$ disjoint children $\{S_k^j\} (0\le j< N)$ and posts for each child a compact commitment (subgraph inclusion Merkle proof and I/O hashes).

\Item{} \textbf{Selection (first offending child).} The challenger recomputes the same partitions locally and designates the first offending child $S_k^{j^\star}$ in topological order over $V$ guided by empirical error percentile thresholds, thereby implicitly agreeing on all earlier children and on the inputs to $S_k^{j^\star}$.

\Item{} \textbf{Advance.} The coordinator backend updates the dispute game state: sets $S_{k+1} \leftarrow S_k^{j^\star}$ and increments $k$, and enforces per-round timeouts. Failure to act within the timeout loses that round.

The game repeats until $|S_k|=1$, at which point the dispute has been localized to a single operator and Phase~3 is invoked. The number of rounds is $O(\log_N |V|)$.

\Paragraph{Phase 3: Single-operator adjudication.} At the leaf, both parties agree on the operator type $v^*$ and the inputs $a$. The \emph{challenger} then tests for fraud using one of two methods: (i) a \textit{theoretical-bound} check, where a verifiable VM computes element-wise FP error bounds (Sec.~\ref{sec:theoretical-bound-computation}): If the proposer's claimed outputs fall outside these bounds for any element, the coordinator confiscates \(\sigma_P\) and rewards the challenger. (ii) When the bound is inconclusive (the claim lies within the theoretical error bound), a \textit{committee} check, where multiple independent runners execute $(v^*,a)$ and vote using the pre-calibrated empirical error percentile thresholds: this test is \emph{much tighter} but more costly.

\vspace{-0.3em}
\section{Floating-point rounding error bounds and empirical thresholds for neural networks}
\label{sec:error-bounds}

In this section, we compute per-operator theoretical floating-point error bounds and calibrate empirical error percentile thresholds, enabling tolerance-aware per-operator verification across devices.

\subsection{Computing theoretical error bound}
\label{sec:theoretical-bound-computation}

This section derives first-order operator-local rounding-error bounds from the IEEE-754 model and shows how to compute them efficiently via FX-based co-execution templates.

\Paragraph{Instrumentation, scope and FP model.}
We instrument PyTorch at operator granularity and compute \emph{operator-local} theoretical floating-point error bounds. Each operator $v$ returns both its computed tensor $\hat{y}_v$ and a same-shape bound $\tau_v^{\text{theo}}$, certifying $y_v \in [\,\hat{y}_v-\tau_v^{\text{theo}},\; \hat{y}_v+\tau_v^{\text{theo}}\,]$.

Crucially, we \emph{do not} propagate bounds across operators (network-level propagation is impractical for deep models); we only account for (i) error \emph{propagated within} operator’s own sub-steps and (ii) \emph{fresh rounding errors} from those steps.

We adopt the standard floating-point model as follows: for  $\circ\in\{+,-,\times,/\}$, $\operatorname{fl}(x\circ y)=(x\circ y)(1+\delta)$ with $|\delta|\le u$ ($u$ is unit roundoff). For library intrinsics we use vendor-stated maximum-ULP (Unit in the Last Place) errors for CUDA math functions~\cite{nvidia_cuda_programming_guide}.
For reductions composed of $k$ basic ops, we support both worst-case deterministic ~\cite{higham2002accuracy,goldberg1991every} and relatively tighter probabilistic~\cite{madry2018towards, higham2019new} bounds. A canonical example is sequential summation $s=\sum_{i=1}^n x_i$ ($k=n-1$):
\begingroup
\setlength{\abovedisplayskip}{2pt}      
\setlength{\belowdisplayskip}{2pt}       
\setlength{\abovedisplayshortskip}{0pt} 
\setlength{\belowdisplayshortskip}{3pt}
\begin{gather*}
\text{deterministic: }\;
|\widehat{s}-s|\le \gamma_{n-1}\sum_{i=1}^n|x_i|,\quad
\gamma_k\triangleq \frac{ku}{1-ku}; \\[-2ex]
\text{probabilistic: }\;
\Pr\!\Big(|\widehat{s}-s|\le \widetilde{\gamma}_{n-1}(\lambda)\sum_{i=1}^n|x_i|\Big)\ge
1-2e^{-\frac{\lambda^2(1-u)^2}{2}},
\end{gather*}
\endgroup
with $\widetilde{\gamma}_k(\lambda)\triangleq \exp\!\big(\lambda\sqrt{k}u+\frac{ku^2}{1-u}\big)-1$.
We use $\lambda=4$, giving $\widetilde{\gamma}_k(4)\approx 4u\sqrt{k}$ at $\ge 99.93\%$ confidence; more details are shown in Appendix.

\Paragraph{FX-based co-execution and operator templates.}
We extend PyTorch FX~\cite{pytorch-fx} interpreter to co-execute values and bounds. Each traced operator is lowered to a short sequence of primitives (adds/multiplications/reductions/intrinsics). For a primitive $f(x_1, \dots, x_m)$ with current inputs $\{x_i\}$ and already-accumulated intra-operator input errors $\{\varepsilon_{x_i}\}$, we use a first-order sensitivity envelope
\begingroup
\setlength{\abovedisplayskip}{2pt}      
\setlength{\belowdisplayskip}{2pt}       
\setlength{\abovedisplayshortskip}{0pt} 
\setlength{\belowdisplayshortskip}{3pt}
$$
\underbrace{\varepsilon^{\text{prop}}_{f}}_{\text{propagated}} \;\le\; \sum_{i=1}^m \Big|\frac{\partial f}{\partial x_i}\Big|\,\varepsilon_{x_i},
\qquad
\underbrace{\varepsilon^{\text{rnd}}_{f}}_{\text{fresh rounding}} \;\le\; u\,|\hat{f}|,
$$
\endgroup
and set $\varepsilon_f=\varepsilon^{\text{prop}}_{f}+\varepsilon^{\text{rnd}}_{f}$.
Reductions (sum/mean/matmul/conv) use $\gamma_k$ or $\widetilde{\gamma}_k$ with the appropriate $k$ per output element for the fresh rounding term; pure data movement (e.g., \texttt{view}/\texttt{cat}/\texttt{gather}) contributes no FP error. 

\Paragraph{Minimal example: \texttt{softmax}.}
For a vector $x$ along dimension $d$, we compute the \texttt{softmax} operator $\hat{y}$ as:
$$
m=\max(x),\; z=x-m,\; e=\exp(z),\; S=\textstyle\sum\nolimits_j e_j,\; \hat{y}_i=e_i/S.
$$

The per-step bounds (all elementwise) separate propagated vs. fresh rounding (first-order worst-case model):

$$
\begin{aligned}
\varepsilon_z &\le u\,(|x|+|m|), \qquad  
\varepsilon_e \le |e|\,\varepsilon_z \;+\; 2u\,|e|, \\
\varepsilon_S &\le \widetilde{\gamma}_{n-1}\!\sum_j |e_j|+(\widetilde{\gamma}_{n-1}+1)\sum_j|\varepsilon_{e_j}|, \\
\varepsilon_{\hat{y}_i} &\le \frac{\varepsilon_{e_i}}{|S|} \;+\; \frac{|e_i|\,\varepsilon_S}{|S|^2} \;+\; u\,|\hat{y}_i|.
\end{aligned}
$$

Here the derivative terms $|\partial \exp/\partial z|=|e|$ and $|\partial (e/S)/\partial(\cdot)|$ generate the \emph{propagated} components, while $u|\cdot|$ terms are \emph{fresh} rounding. Analogous templates cover other 35 common operators (\texttt{LayerNorm}, \texttt{MatMul}, \texttt{Conv2d}, etc.); we ignore the additional numerical errors introduced while computing these error bounds themselves.

\begin{figure}
    \centering
    \includegraphics[width=\linewidth]{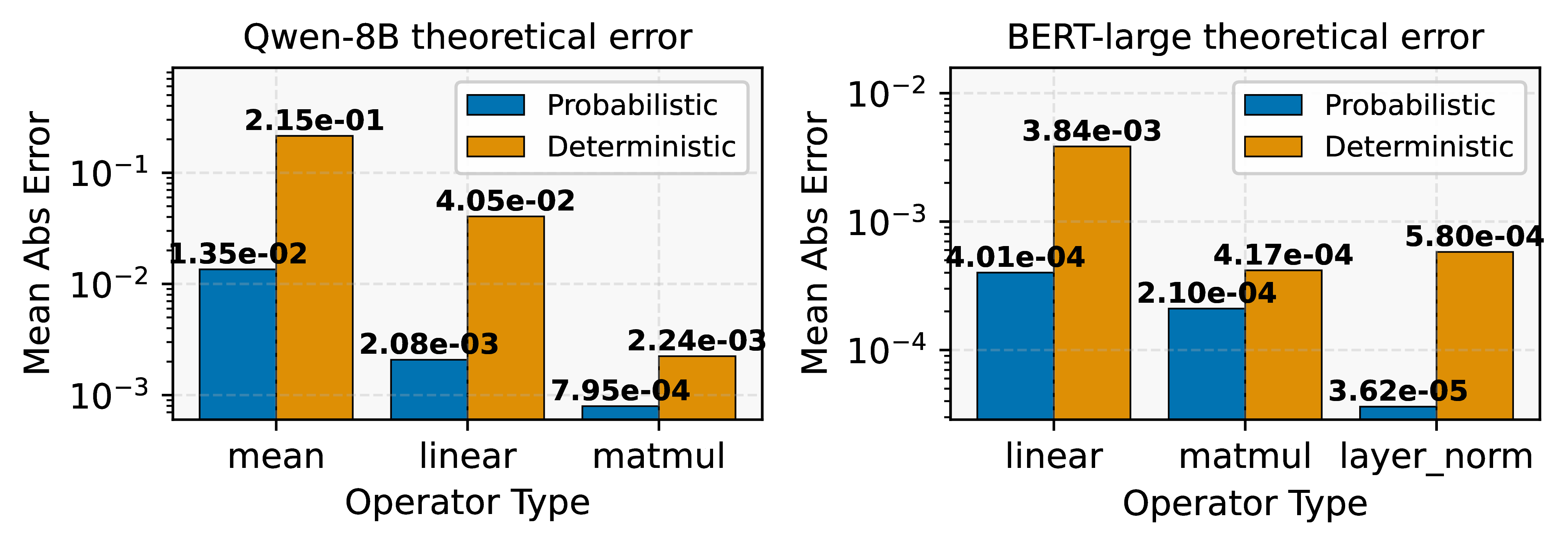}
    \vspace{-25pt}
    \caption{Comparison between deterministic and probabilistic theoretical error bounds for different operator types in Qwen3-8B and BERT-large. }
    \label{fig:theoretical-deterministic-and-probabilistic-comparison}
    \Description{A graph comparing deterministic and probabilistic theoretical error bounds for different operator types in Qwen3-8B and BERT-large. The x-axis represents the node types, while the y-axis the mean of absolute error. The graph shows that probabilistic bounds are significantly tighter than deterministic bounds. }
\end{figure}

\Paragraph{Putting it together.}
Our library of operator templates yields $\tau_v^{\text{theo}}$ cheaply during a single traced run. These per-operator bounds are later used for single-operator verification. Fig.~\ref{fig:theoretical-deterministic-and-probabilistic-comparison} contrasts deterministic $\gamma_{k}$ with probabilistic $\widetilde{\gamma}_{k}$ across representative operators; the probabilistic bounds are markedly tighter, especially at large $k$. These bounds provide a cheap, hardware-agnostic ``ground-truth'' leaf check that doesn't rely on committee honesty. Many real-world misbehaviors (e.g., model swaps, undeclared quantization) typically deviate far outside theoretical bounds and can be rejected quickly. 

However, worst-case $\gamma_{k}$ and even $\widetilde{\gamma}_{k}$ remain conservative for real ML workloads where mixed signs induce cancellation; empirically, observed cross-device discrepancies are often orders of magnitude smaller. We therefore pre-calibrate per-operator error percentile profiles on representative inputs to construct empirical thresholds and use them alongside the theoretical bounds.

\subsection{Calibrating empirical error percentile threshold}
\label{sec:empirical-error-percentile-threshold-calibration}

In this section, we calibrate device-agnostic absolute/relative error percentile profiles for each operator across inputs and hardware. We then validate their stability and inflate them into practical thresholds with a safety margin. 

The dispute protocol narrows disagreements from the full computation graph to individual operators by steering a N-way partition search with empirically calibrated error percentile thresholds. Across inputs, intermediate tensor shapes may differ, so for feasible comparisons, we calibrate per-operator error distributions rather than using element-wise bounds. We describe our offline calibration in this section. 

\Paragraph{Cross-device calibration procedure.} For a traced model $G = (V, E)$, we sample a representative dataset $\mathcal{D} = \{x_1, \ldots, \allowbreak x_m\}$ and evaluate it on multiple GPU devices $\mathcal{H} = \{h_1, \ldots, h_k\}$. For operator $v_i \in V$ and input $x_s$, let $y_i^{(h_j, x_s)} \in \mathbb{R}^{d_i}$ denote the output tensor of $v_i$ on device $h_j$, where $d_i$ is the tensor dimension after the tensor being flattened.

For each device pair $(h_j, h_k)$, we form element-wise absolute and relative errors:
\begin{align}
    \mathbf{e}_{\mathrm{abs}}^{(i,j,k,s)} &= |y_i^{(h_j, x_s)} - y_i^{(h_k, x_s)}| \in \mathbb{R}^{d_i}, \\
    \mathbf{e}_{\mathrm{rel}}^{(i,j,k,s)} &= \frac{|y_i^{(h_j, x_s)} - y_i^{(h_k, x_s)}|}{|y_i^{(h_j, x_s)}| + \epsilon} \in \mathbb{R}^{d_i},
\end{align}
with a small $\epsilon > 0$ to avoid division by zero.

\Paragraph{Percentile profiles.} For each $(i, j, k, s)$, we compute the \emph{percentile-value vector} over tensor elements at percentiles $p \in P = \{1, 5, 10, 15, \ldots, 85, 90, 95, 99, 100\}$:
\begin{align}
    \mathcal{P}_{\mathrm{abs}}^{(i,j,k,s)}(p) &= \text{Percentile}_p(\mathbf{e}_{abs}^{(i,j,k,s)}), \\
    \mathcal{P}_{\mathrm{rel}}^{(i,j,k,s)}(p) &= \text{Percentile}_p(\mathbf{e}_{rel}^{(i,j,k,s)}),
\end{align}
where $\mathcal{P}(\cdot)$ denotes the vector of percentile profiles (error distribution) indexed by $p\in P$, and is computed over all $d_i$ elements of the error tensor. We obtain conservative, device- and input-agnostic operator-wise profiles by a max-envelope across $j,k,s$:

\begingroup
\setlength{\abovedisplayskip}{-4pt}  
\setlength{\belowdisplayskip}{1pt}
\begin{align}
    \mathcal{P}_{\mathrm{abs}}^{(i)}(p) &= \max_{j,k} \left\{ \{\mathcal{P}_{\mathrm{abs}}^{(i,j,k,s)}(p)\}_{s=1}^m\right\}, \\
    \mathcal{P}_{\mathrm{rel}}^{(i)}(p) &= \max_{j,k} \left\{ \{\mathcal{P}_{\mathrm{rel}}^{(i,j,k,s)}(p)\}_{s=1}^m\right\}.
\end{align}
\endgroup

\begin{table}
\centering
\scriptsize
\caption{Stability metrics at selected percentiles (p30, p50, p70) for three models. All values are normalized by median.}
\label{tab:stability-metrics}
\setlength{\tabcolsep}{5.5pt}
\begin{tabular}{llcccccccc}
\toprule
\multirow{2}{*}{Model} & \multirow{2}{*}{$p$} & \multicolumn{2}{c}{SupNorm} & \multicolumn{2}{c}{Jackknife} & \multicolumn{2}{c}{TailAdj} & \multicolumn{2}{c}{RollSD} \\
\cmidrule(lr){3-4}\cmidrule(lr){5-6}\cmidrule(lr){7-8}\cmidrule(lr){9-10}
 &  & @50 & @90 & @50 & @90 & @50 & @90 & @50 & @90 \\
\midrule
\multirow{3}{*}{Qwen}
 & 30 & 0.00 & 0.03 & 0.00 & 0.02 & 0.00 & 0.03 & 0.05 & 0.11 \\
 & 50 & 0.00 & 0.02 & 0.00 & 0.02 & 0.00 & 0.03 & 0.05 & 0.11 \\
 & 70 & 0.00 & 0.02 & 0.00 & 0.02 & 0.01 & 0.03 & 0.05 & 0.10 \\
\midrule
\multirow{3}{*}{BERT}
 & 30 & 0.00 & 0.04 & 0.00 & 0.00 & 0.00 & 0.03 & 0.03 & 0.10 \\
 & 50 & 0.00 & 0.04 & 0.00 & 0.00 & 0.00 & 0.03 & 0.03 & 0.09 \\
 & 70 & 0.00 & 0.04 & 0.00 & 0.01 & 0.00 & 0.02 & 0.03 & 0.08 \\
\midrule
\multirow{3}{*}{ResNet}
 & 30 & 0.00 & 0.02 & 0.00 & 0.00 & 0.00 & 0.03 & 0.00 & 0.10 \\
 & 50 & 0.00 & 0.04 & 0.00 & 0.01 & 0.00 & 0.03 & 0.00 & 0.09 \\
 & 70 & 0.00 & 0.05 & 0.00 & 0.01 & 0.00 & 0.03 & 0.00 & 0.10 \\
\bottomrule
\end{tabular}
\end{table}

\Paragraph{Statistical validation.} We sampled 50 inputs for each model to collect the percentile profiles of each operator. We assess per-operator stability with four metrics: SupNorm (drift), Jackknife (leave-one-out influence), TailAdj (recent tail change), and RollSD computed over running medians (window $W=10$) and summarized at the 50th/90th percentiles across operators, normalized by each metric's median (Table~\ref{tab:stability-metrics}; details in Sec.~\ref{sec:statistical-validation}). Across three models and $p\!\in\!\{30,50,70\}$, central tendencies are $\approx0$, and 90th-percentile bounds are tight (SupNorm $\le0.05$, Jackknife $\le0.02$, TailAdj $\le0.03$); RollSD is modestly higher (up to $0.11$). These indicate near-stationary operator estimates and support the reliability of our empirical error profiles. The metric details are in supplementary materials.

\Paragraph{Threshold construction.} To account for calibration uncertainty and provide a safety margin, we apply a multiplicative factor $\alpha = 3$ to every percentile value for each $v_i \in V$:
\begin{align}
    \tau_{\mathrm{abs}}^{(i)}(p) = \alpha \cdot \mathcal{P}_{\mathrm{abs}}^{(i)}(p), \qquad
    \tau_{\mathrm{rel}}^{(i)}(p) = \alpha \cdot \mathcal{P}_{\mathrm{rel}}^{(i)}(p).
\end{align}
The set $\{\tau_{\mathrm{abs}}^{(i)}, \tau_{\mathrm{rel}}^{(i)}\}_{i=1}^{|V|}$ is bundled with the model commitment $(r_w, r_g)$ and is included in the model commitment recorded by the coordinator.

Taken together, theoretical bounds provide sound, portable bounds, while empirical profiles yield tight, model-specific guidance. This combination enables well-posed, tolerance-aware verification that respects IEEE-754 nondeterminism without sacrificing production performance.
\section{Threat model and attack design}
\label{sec:adversary-model}

This section evaluates adversarial proposer behavior under the protocol's numerical admissible sets; it does not attempt to model all deployment risks (e.g., coordinator compromise).
We then describe one of the attack methods used to probe these interfaces under their respective conditions and, finally, demonstrate the robustness of the error bounds.

\subsection{Adversary and objectives}

\Paragraph{Adversary capabilities.} 
In this section, we consider the case where only the \emph{proposer} is adversarial; all other parties follow the protocol. The adversary is \emph{white-box}: they know model commitment, input, canonical topological order, empirical error percentile thresholds, and the dispute-partition policy. The adversary may alter intermediate tensors during the computation but \emph{cannot} change weights or graph topology without violating Merkle commitments, nor influence committee sampling or cryptography.

\Paragraph{Security objectives.}
We evaluate:
\emph{(i) Soundness}: deviations that beyond committed bounds are detected (by empirical thresholds or theoretical bounds);
\emph{(ii) Non-accumulation}: admissible per-operator deviations do not compose into large task-level error;
\emph{(iii) Low false positives}: honest executions rarely trigger disputes under empirical thresholds.

\begin{figure}
  \centering
  \includegraphics[width=0.8\linewidth]{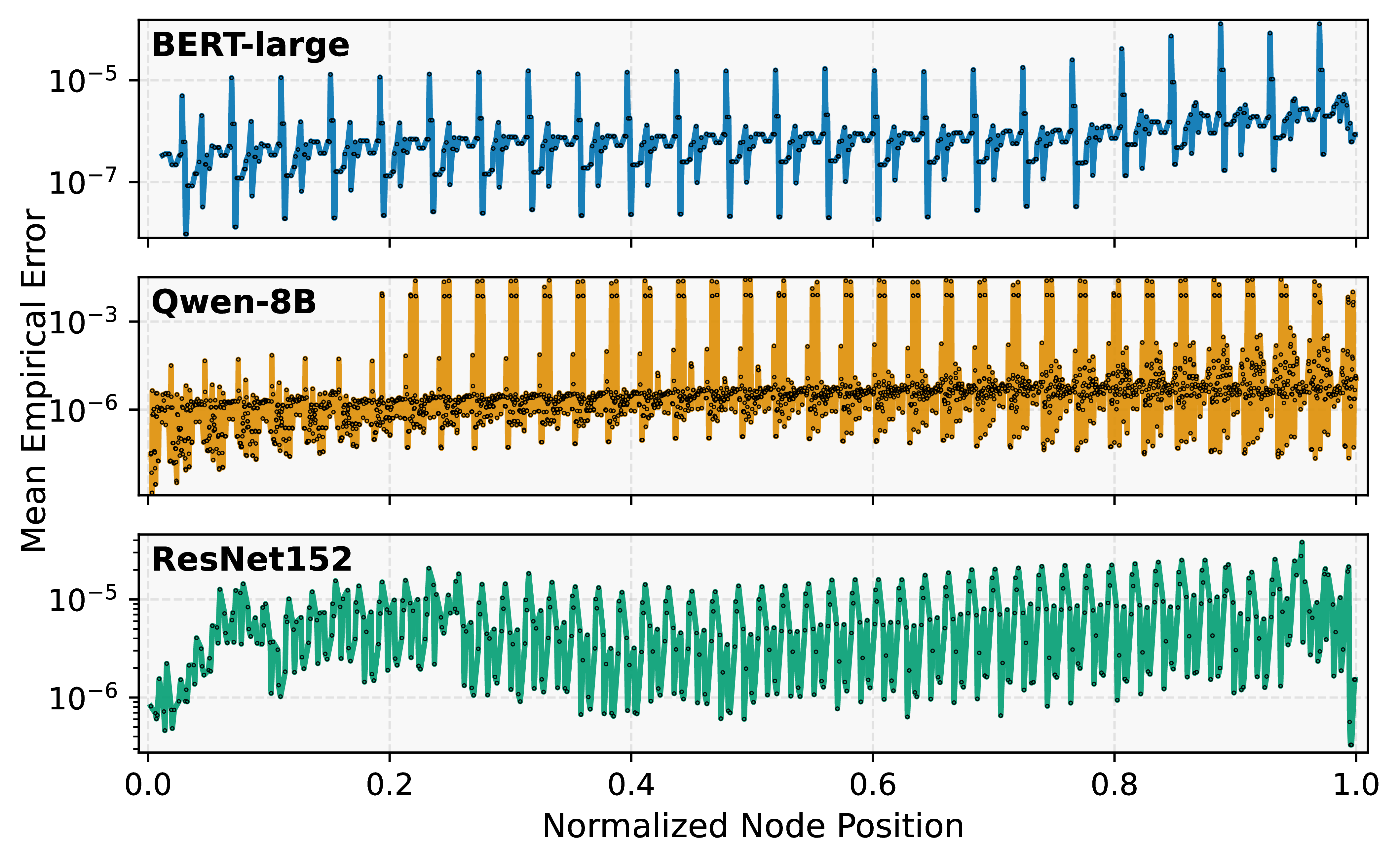}
  \vspace{-15pt}
  \caption{Mean empirical errors vs.\ normalized operator position (log-scale y-axis). The x-axis follows the canonical topological order. Profiles remain essentially flat with localized spikes, indicating limited room for attack.}
  \label{fig:empirical-error-evolution}
  \Description{Line plots showing mean empirical errors versus normalized operator position for multiple models on a log scale. Profiles remain flat with localized spikes.}
  \vspace{-15pt}
\end{figure}

\subsection{Attack surface and empirical headroom}

The attack surface is numerical: at selected $v$, the proposer injects additive perturbations $h_v \leftarrow h_v + \Delta_v$ to change the final decision while remaining within challenger's admissible sets. To quantify headroom, we trace mean cross-device error per operator and plot it against normalized depth (Fig.~\ref{fig:empirical-error-evolution}). Across \texttt{BERT-large}, \texttt{Qwen3-8B}, and \texttt{ResNet-152}, errors are essentially flat with localized spikes: we observe no systematic accumulation of error as depth increases. Typical magnitudes are $10^{-6}$-$10^{-5}$. Thus, flipping outputs while staying admissible requires coordinating many extremely small deviations; over-allocation will be isolated by the dispute game.

\subsection{Evasion strategies}
\label{sec:attack-families}
In this section, we define two evasion strategies aligned with search-time and leaf-time verification checks: the adversary chooses $\{\Delta_v\}$ to change the final decision while remaining within the verifier's admissible sets.

\Paragraph{Empirical-threshold evasion (search time).}
For operator $v_i$, calibration commits percentile-cap pairs $\{(p_k,\tau_{\mathrm{abs}}^{(i)}(p_k))\}_{k=1}^K$ and relative caps $\tau_{\mathrm{rel}}^{(i)}(\cdot)$. Define the nondecreasing cap curve $C_i:[0,100]\!\to\!\mathbb{R}_{\ge 0}$ by linear interpolation through $(0,0)$, $(p_k,\tau_{\mathrm{abs}}^{(i)}(p_k))$, and $(100,\tau_{\mathrm{abs}}^{(i)}(100))$. The feasible set is
\begingroup
\setlength{\abovedisplayskip}{3pt}  
\setlength{\belowdisplayskip}{3pt}
\begin{equation}
\mathcal{F}^{\text{emp}}_{v_i} \triangleq \Big\{\Delta:\ Q_{|\,\Delta\,|}(r)\le C_i(r),\ \forall r\in[0,100]\Big\},
\end{equation}
\endgroup
where $Q_{|\,\Delta\,|}$ is the empirical quantile function of $|\Delta|$. Evasion requires the observed proposer–challenger discrepancy at each inspected operator to stay within $\mathcal{F}^{\text{emp}}_{v_i}$  for given ${p_k}$.

\Paragraph{Theoretical-bound evasion (leaf check).}
At the leaf, the challenger either runs the \emph{theoretical} element-wise IEEE-754 bound or invokes a small committee vote. Here we focus on the attack against the theoretical bound check (latter is the empirical threshold check). For operator $v^\star$ with bound $\tau^{\text{theo}}_{v^\star}$, the feasible set is
\begingroup
\setlength{\abovedisplayskip}{3pt}  
\setlength{\belowdisplayskip}{3pt}
\begin{equation}
\mathcal{F}^{\text{theo}}_{v^\star} \triangleq \big\{\Delta:\ |\Delta|\le \tau^{\text{theo}}_{v^\star}\ \text{element-wise}\big\}.
\end{equation}
\endgroup
Because these worst-case bounds are typically looser than empirical error thresholds, the protocol falls back to a committee vote against empirical thresholds when the theoretical check is inconclusive. 

\begin{table*}
  \centering
  \scriptsize
  \caption{\textbf{Bucketed attack outcomes under threshold scaling.}
  Each cell reports \emph{ASR (\%)} and mean $\Delta m_{\text{fail}}$ ($\delta_{\text{fail}}$). Buckets group targets by their logit margin percentile. Scale $\alpha$ multiplies the error bounds (larger is looser). \emph{False Positive (\%)} is the honest-run dispute rate at the same $\alpha$. (For theoretical bound attack, d is deterministic bound; p is probabilistic bound)}
  \label{tab:main-threshold-scaling-full}
  \vspace{-12pt}
  \setlength{\tabcolsep}{3pt}
  \newcommand{\twohdr}[1]{\multicolumn{2}{c}{#1}}
  \newcommand{\paircols}{ASR (\%) & $\overline{\Delta m}_{\text{fail}}$($\overline{\delta}_{\text{fail}}$)}
  \definecolor{asrgreen}{RGB}{227,252,227}
  \definecolor{asrred}{RGB}{255,235,238} 
  \newcommand{\asrzero}[1]{\multicolumn{1}{>{\columncolor{asrgreen}}c}{\strut #1}}
  \newcommand{\asrpos}[1]{\multicolumn{1}{>{\columncolor{asrred}}c}{\strut #1}}
  
  \begin{tabular}{llc *{5}{cc} c}
  \toprule
  \multirow{2}{*}{Model} & \multirow{2}{*}{Bound check} & \multirow{2}{*}{Scale $\alpha$} &
  \twohdr{0--20\%} &
  \twohdr{20--40\%} &
  \twohdr{40--60\%} &
  \twohdr{60--80\%} &
  \twohdr{80--100\%} &
  \multirow{2}{*}{\parbox{1.5cm}{\centering False\\ Positive (\%)}}
  \\
  \cmidrule(lr){4-5}\cmidrule(lr){6-7}\cmidrule(lr){8-9}\cmidrule(lr){10-11}\cmidrule(lr){12-13}
  & & & \paircols & \paircols & \paircols & \paircols & \paircols & \\
  \midrule
  
  \multirow{6}{*}{BERT-large}
    & \multirow{3}{*}{Empirical}  & $\times$1  & \asrzero{0.0}&0.02(0.3\%)& \asrzero{0.0}&0.02(0.2\%)& \asrzero{0.0}&0.02(0.2\%)& \asrzero{0.0}&0.02(0.2\%)& \asrzero{0.0}&0.02(0.2\%)& 0 \\
    &                             & $\times$2  & \asrzero{0.0}&0.04(0.5\%)& \asrzero{0.0}&0.04(0.4\%)& \asrzero{0.0}&0.04(0.4\%)& \asrzero{0.0}&0.04(0.4\%)& \asrzero{0.0}&0.03(0.3\%)& 0 \\
    &                             & $\times$3  & \asrzero{0.0}&0.06(0.8\%)& \asrzero{0.0}&0.06(0.6\%)& \asrzero{0.0}&0.06(0.6\%)& \asrzero{0.0}&0.06(0.6\%)& \asrzero{0.0}&0.05(0.5\%)& 0 \\
  \cmidrule(lr){2-14}
    & \multirow{3}{*}{Theo (d/p)} & $\times$1(d)  &\asrpos{58.6}&8.34(94\%)&\asrpos{32.6}&8.32(88\%)&\asrpos{26.4}&8.56(88\%)&\asrpos{20.8}&8.73(87\%)&\asrpos{14.6}&8.73(84\%)& - \\
    &                              & $\times$1(p)  & \asrzero{0.0}&0.71(8.9\%)& \asrzero{0.0}&0.62(6.9\%)& \asrzero{0.0}&0.64(6.9\%)& \asrzero{0.0}&0.61(6.3\%)& \asrzero{0.0}&0.56(5.6\%)& - \\
    &                              & $\times$0.5(p)& \asrzero{0.0}&0.32(4.1\%)& \asrzero{0.0}&0.28(3.1\%)& \asrzero{0.0}&0.28(3.0\%)& \asrzero{0.0}&0.27(2.8\%)& \asrzero{0.0}&0.24(2.4\%)& - \\
  \midrule
  
  \multirow{6}{*}{ResNet-152}
    & \multirow{3}{*}{Empirical}  & $\times$1  & \asrzero{0.0} & 0.31(4.0\%) & \asrzero{0.0} & 0.30(3.8\%) & \asrzero{0.0} & 0.30(3.7\%) & \asrzero{0.0} & 0.30(3.6\%) & \asrzero{0.0} & 0.30(3.6\%) & 0 \\
    &                             & $\times$2  & \asrzero{0.0} & 0.62(8.0\%) & \asrzero{0.0} & 0.60(7.4\%) & \asrzero{0.0} & 0.59(7.2\%) & \asrzero{0.0} & 0.60(7.1\%) & \asrzero{0.0} & 0.60(7.0\%) & 0 \\
    &                             & $\times$3  & \asrzero{0.0} & 0.94(12\%)  & \asrzero{0.0} & 0.91(11\%)  & \asrzero{0.0} & 0.89(11\%)  & \asrzero{0.0} & 0.89(11\%)  & \asrzero{0.0} & 0.90(10\%) & 0 \\
  \cmidrule(lr){2-14}
    & \multirow{3}{*}{Theo (d/p)} & $\times$1(d)  &\asrzero{0.0}&2.67(34\%)&\asrzero{0.0}&2.54(31\%)&\asrzero{0.0}&2.50(30\%)&\asrzero{0.0}&2.49(29\%)&\asrzero{0.0}&2.46(29\%)& - \\
    &                              & $\times$1(p)  & \asrzero{0.0} & 0.50(6.7\%) & \asrzero{0.0} & 0.48(6.1\%) & \asrzero{0.0} & 0.48(6.0\%) & \asrzero{0.0} & 0.48(5.9\%) & \asrzero{0.0} & 0.48(5.7\%) & - \\
    &                              & $\times$0.5(p)& \asrzero{0.0} & 0.25(3.3\%) & \asrzero{0.0} & 0.24(3.0\%) & \asrzero{0.0} & 0.24(3.0\%) & \asrzero{0.0} & 0.24(2.9\%) & \asrzero{0.0} & 0.24(2.8\%) & - \\
  \midrule
  
  \multirow{6}{*}{Qwen3-8B}
    & \multirow{3}{*}{Empirical}  & $\times$1  & \asrzero{0.0} & 0.77(4.2\%) & \asrzero{0.0} & 0.79(3.8\%) & \asrzero{0.0} & 0.81(3.6\%) & \asrzero{0.0} & 0.82(3.5\%) & \asrzero{0.0} & 0.87(3.4\%) & 0 \\
    &                             & $\times$2  & \asrzero{0.0} & 1.45(7.9\%) & \asrzero{0.0} & 1.46(7.0\%) & \asrzero{0.0} & 1.51(6.7\%) & \asrzero{0.0} & 1.54(6.5\%) & \asrzero{0.0} & 1.60(6.3\%) & 0 \\
    &                             & $\times$3  & \asrzero{0.0} & 2.05(11\%)  & \asrzero{0.0} & 2.08(9.8\%)  & \asrzero{0.0} & 2.14(9.5\%)  & \asrzero{0.0} & 2.18(9.2\%)  & \asrzero{0.0} & 2.26(8.9\%) & 0 \\
  \cmidrule(lr){2-14}
    & \multirow{2}{*}{Theo (d/p)} & $\times$1(d)  & \asrpos{12.6}&18.9(98\%)&\asrpos{9.2}&19.4(96\%)&\asrpos{6.4}&19.7(95\%)&\asrpos{9.2}&20.5(94\%)&\asrpos{8.2}&21.2(89\%) & - \\
    &                              & $\times$1(p)  & \asrpos{2.4} & 7.31(40\%) & \asrpos{2.4} & 7.46(36\%) & \asrpos{1} & 7.76(35\%) & \asrpos{1.2} & 8.05(34\%) & \asrpos{0.6} & 8.35(33\%) & - \\
    &                              & $\times$0.5(p)& \asrpos{0.6} & 5.87(31\%) & \asrpos{0.2} & 5.89(28\%) & \asrpos{0.8} & 6.02(27\%) & \asrpos{0.2} & 6.13(26\%) & \asrpos{0.6} & 6.36(25\%) & - \\
  \bottomrule
  \end{tabular}
  \end{table*}

\subsection{Optimization-based attack method}
\label{sec:attack-method}

We implement projected-gradient attack~\cite{madry2018towards} over $\{\Delta_v\}$ to flip the predicted class/token while staying within a chosen admissible set.

\Paragraph{Attack objective.}
Let $z\in\mathbb{R}^C$ be the logits at model's output, and we maximize the logit margin:
\begin{equation}
L_{\text{margin}}(z) \;=\; z_{c_2}-z_{c_1},
\end{equation}
where $c_1=\arg\max_c z_c$ and $c_2$ is the attack's target class. It succeeds once $L_{\text{margin}}>0$.

\Paragraph{PGD updates and projections.}
At iteration $t$, compute $g_v^t=\nabla_{\Delta_v} L_{\text{margin}}$ through the traced graph, apply Adam with $(\beta_1,\beta_2,\varepsilon)=(0.9,0.999,10^{-8})$ and a per-operator stepsize $\eta_v$ (default: $1/4$ median of its error bound), to get a tentative $\widetilde\Delta_v^{\,t+1}$. We then project onto the relevant feasible set:

\emph{(i) Projection under theoretical bounds.}
This projection is simply element-wise clipping ($\tau_v$ is computed at runtime using this operator's input):
\begin{equation}
\Delta_v^{t+1}=\operatorname{clip}\!\left(\widetilde\Delta_v^{\,t+1},-\tau_v,\tau_v\right).
\end{equation}
\emph{(ii) Projection under empirical thresholds.}
Let $a=\bigl|\mathrm{vec}(\widetilde\Delta_v^{\,t+1})\bigr|\in\mathbb{R}^n$, $s=\mathrm{sign}(\mathrm{vec}(\widetilde\Delta_v^{\,t+1}))$, and let $\sigma$ sort $a$ increasingly. 
For ranks $r_k=(k-\tfrac{1}{2})/n$, define caps $c_k=C_v(r_k)$ and enforce monotonicity via $c_k\leftarrow\max(c_k,c_{k-1})$. 
Clip order statistics $a^\star_{\sigma(k)}=\min\{a_{\sigma(k)},c_k\}$, then restore sign/shape:
\begin{equation}
\Delta_v^{t+1}=\mathrm{reshape} \;\!\big(s\odot a^\star\big).
\end{equation}
This realizes the projection onto $\mathcal{F}^{\text{emp}}_v$. The dominant cost is the sort: $O(n\log n)$ per operator; all steps are vectorized.

\subsection{Attack evaluation}
\label{sec:attack-evaluation}

In this section, we measure attack success, progress, and false positives across models under threshold scaling. Experiments are conducted on \texttt{ResNet-152}, \texttt{BERT-large}, and \texttt{Qwen3-8B}.

\Paragraph{Datasets and hyperparameters.}
For \texttt{BERT-large}~\cite{devlin2019bert} we use DBpedia~\cite{auer2007dbpedia}; for \texttt{Qwen3-8B}~\cite{qwen3technicalreport} we use C4~\cite{2019t5}, feeding the first 10\% of each sequence and targeting the next token; for \texttt{ResNet-152}~\cite{he2016deep} we use ImageNet~\cite{deng2009imagenet} validation set. PGD/Adam and projection operators follow Sec.~\ref{sec:attack-method}. Early stopping triggers when the last 10 updates satisfy $|m_t-m_{t-1}|<10^{-3}|m_0|$ and $|m_t|<10^{-3}|m_0|$ ($m_t$ is the logit margin at time step $t$). We use 500 samples (2500 attack targets) for the theoretical-bound attack and 250 (1250 attack targets) for the empirical-threshold attack to balance statistical precision and computational cost.

\Paragraph{Metrics and bucketing.} We report (i) \emph{attack success rate} (ASR), the fraction of attacks that successfully flip the model's prediction to the target class while staying under the chosen bound, and (ii) the mean logit margin change on \emph{attack failed} runs, $\overline{\Delta m}_{\text{fail}}$, measuring how close failed attacks came to success, and $\overline{\delta}_{\text{fail}}$ measures the normalized margin change to the target margin. For example, let $m_0 = z_{c_1} - z_{c_2} > 0$ be the logit margin between the original predicted token $c_1$ and the target token $c_2$ before attack begins. After the attack process, the new logit gap is $m' = z'_{c_1} - z'_{c_2}$. Then, $\Delta m = m_0 - m'$ and $\delta = \frac{\Delta m}{m_0}$. To study the perturbation for different levels of token / classes, for each model input, we bucket and sample examples by the target's logit margin in the original prediction (lower means smaller margins): $[0,20\%], [20\%,40\%], [40\%,60\%], [60\%,80\%], [80\%,100\%]$. We randomly sampled one class from each bucket to attack. We also scale bounds by a factor $\alpha$ to probe sensitivity: $\alpha>1$ loosens empirical error percentile thresholds; $\alpha<1$ tightens theoretical error bounds (diagnostic only).
For theoretical error bounds we show \emph{d} = deterministic theoretical bounds and \emph{p} = probabilistic theoretical bounds (Sec.~\ref{sec:error-bounds}). This comparison also shows the reason for selecting probabilistic bounds in practice. \emph{False Positive} (FP) is measured by running the full pipeline on honest executions across 1k inputs sampled from the dataset and reporting the rate at which the pipeline raises a dispute against honest computations.

\Paragraph{Results and analysis.} Table~\ref{tab:main-threshold-scaling-full} summarizes the attack outcomes. We can conclude the following:

\emph{(1) Empirical-threshold is robust across models and all target attack settings.}
With empirical error percentile thresholds, ASR remains $0.0\%$ for all three models even when bounds are loosened by $3\times$ ($\alpha=3$). The normalized progress on failed attack is minimal, with $\overline{\delta}_{\text{fail}} < 1\%$ for \texttt{BERT-large} and $\overline{\delta}_{\text{fail}} < 10\%$ for \texttt{ResNet-152} and \texttt{Qwen3-8B}. It demonstrates the effectiveness of our empirical threshold verification.

\emph{(2) Theoretical bound check leaves a small attack window for LLMs.} Under the theoretical element-wise bounds at the leaf, \texttt{BERT-large} and \texttt{ResNet-152} still maintain $0.0\%$ ASR, while \texttt{Qwen3-8B} exhibits low yet non-zero vulnerability: up to $2.4\%$ ASR at $\alpha=1$ and $0.6\%$ ASR at $\alpha=0.5$. The mean progress on failed attacks is significant, with $\overline{\delta}_{\text{fail}}$ up to $40\%$ at $\alpha=1$. These cases align with the intuition that larger models (LLMs) can be sensitive to small logit perturbations; worst-case IEEE-754 theoretical bounds are loose and may not always rule the malicious perturbations out. It also justifies our choice of probabilistic theoretical bounds over deterministic ones, as the latter leaves a much larger attack window. The comparison between (1) and (2) also highlights the advantage of empirical thresholds over theoretical worst-case bounds in practice for a meticulous attacker.

\emph{(3) No false positives with empirical bounds.} Across all models and scaling factors, the full dispute pipeline with empirical bounds never raises a dispute against honest nodes, indicating that our empirical bounds are not only robust against attacks but also practical for real-world deployment.

\begin{figure}
  \centering
  \includegraphics[width=\linewidth]{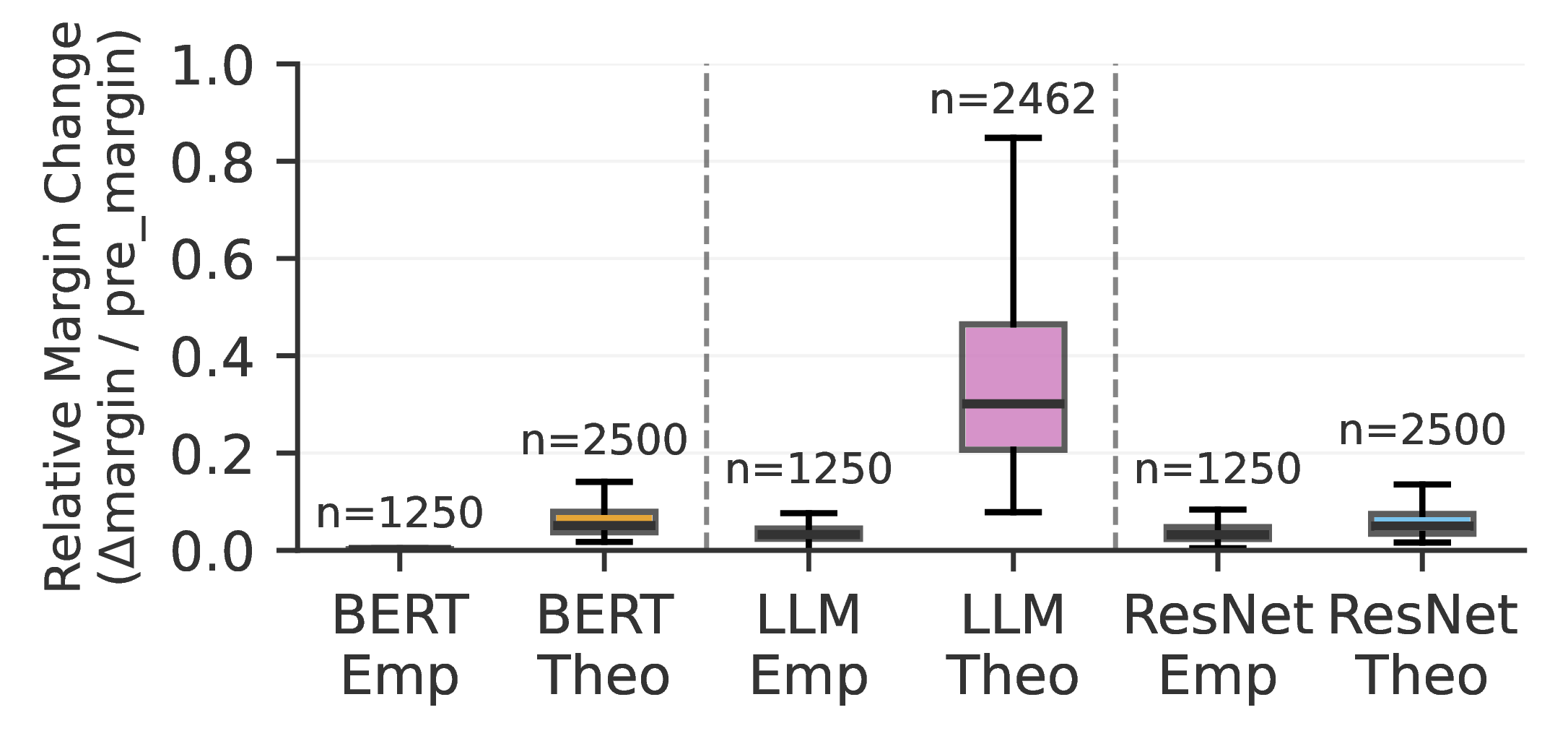}
  \vspace{-22pt}
  \caption{Normalized margin change on failed attacks ($\alpha{=}1$). Empirical thresholds yield near-zero progress across models; theoretical bounds(p) show heavier tails for the LLM. Numbers above boxes show sample sizes.}
  \label{fig:margin-change-boxplot}
  \Description{A box plot showing the normalized margin change on failed attacks with alpha equal to 1. Empirical thresholds yield near-zero progress across models, while theoretical bounds show heavier tails for the LLM. Numbers above boxes indicate sample sizes.}
  \vspace{-22pt}
\end{figure}

\emph{(4) Margin-change distributions corroborate tightness.}
Fig.~\ref{fig:margin-change-boxplot} presents a boxplot of the normalized margin change for the failed cases after the attack. The empirical threshold attacks are all tightly concentrated at ~0.05, indicating that admissible perturbations yield almost no progress towards flipping the decision. In contrast, the theoretical bound attack for \texttt{Qwen3-8B} exhibits a pronounced step-up, indicating that some perturbations can make significant progress towards flipping its decision under the theoretical bounds.

\Paragraph{Key insights.}
Empirical error percentile thresholds keep disputes localized and robust without determinism, while theoretical error bounds give a low-cost, portable fallback at the leaf. The tradeoff is model-dependent: for LLMs, worst-case IEEE-754 bounds can be permissive, motivating a committee vote when tighter adjudication is desired. 
\section{\sys : Optimistic operator-level dispute protocol}
\label{sec:methodology}

In this section, we adopt an \emph{optimistic} design in which a coordinator acts as the adjudicator. A \emph{proposer} commits to the model, inputs, and outputs; if no \emph{challenger} disputes within a window, the result finalizes cheaply. Upon dispute, an interactive localization game narrows disagreement to a single operator guided by the error tolerance, after which the coordinator performs a constant-cost check; the losing party is penalized.

\subsection{Background and design choices.}
Prior verification systems have largely relied on \emph{refereed delegation of computation}, which guarantees correctness by querying multiple untrusted servers and resolving disagreements via a referee~\cite{CanettiRivaRothblumIC13,CRR-CCS11}, and interactive proofs~\cite{GKR08,KRR22}.
Rollup-style verification games (e.g., Truebit, Arbitrum, OP Stack~\cite{teutsch2024scalable,kalodner2018arbitrum,optimismcannonfpvm,optimismfaultproof}) instantiate this over CPU-style VMs via instruction-level bisection. NN on heterogeneous accelerators violates both: GPU kernels cannot be stepped by existing VMs and floating-point non-associativity plus parallel scheduling introduce cross-hardware variability~\cite{pytorch-reproducibility,higham2002accuracy}. 

\sys preserves the optimistic, interactive \emph{structure} but replaces the \emph{units of dispute} and the \emph{base-case check}:
(i) we perform N-way partition on a \emph{Merkle-anchored, operator-granular} NN graph rather than a VM trace; (ii) we guide partition selection via \emph{tolerance-aware} error thresholds calibrated per operator; and (iii) at the leaf we perform either a sound IEEE-754 theoretical-bound check or (when bounds are permissive) a committee vote against empirical error percentile thresholds. This design is new in (a) moving \emph{from instruction-level to operator-level} partition, (b) defining a \emph{deterministic selection rule under non-deterministic floats}, and (c) providing a \emph{constant-cost} adjudication path on the blockchain that respects IEEE semantics without requiring a GPU VM.

\subsection{Operator-granular graph representation and verifiable subgraph extraction}
\label{sec:graph-representation-and-verifiable-subgraph-extraction}

In this section, we trace models to an operator-level dataflow graph and show how to extract, commit to, and verify Merkle-\allowbreak anchored subgraphs.

\Paragraph{Cut sets and subgraph extraction.} Given a candidate subgraph $S$ (a contiguous slice of operators in the execution order), we compute its live-in and live-out frontiers by a linear scan:
\begin{align}
\mathrm{In}(S) \;&=\; \{\, u \notin S \mid \exists\, v \in S: (u \to v)\,\}, 
\\
\mathrm{Out}(S) \;&=\; \{\, w \notin S \mid \exists\, v \in S: (v \to w)\,\}.
\end{align}

We materialize \(S\) as an \texttt{fx.GraphModule} by cloning operators in \(S\), replacing references to \(\mathrm{In}(S)\) with placeholders, and emitting \(\mathrm{Out}(S)\) as graph outputs. Parameters are reused by reference; operator attributes and tensor metadata are preserved.

\Paragraph{Merkle commitments to weights, graph, and interfaces.}
We commit separately to weights and graph structure via Merkle trees whose roots are recorded by the coordinator. We define $\Hsh(\cdot)$ as the SHA-256 hash function.

\emph{Weights.}
We sort \texttt{state\_dict} keys lexicographically and hash each parameter tensor:
\begin{align*}
\mathcal{H}_w(k)=\Hsh\big(\textsf{canon}(\texttt{state\_dict}[k])\big),
\end{align*}
where \(\textsf{canon}(\cdot)\) serializes raw tensor bytes, dtype, shape, and stride, forming the weight Merkle tree \(\mathcal{T}_w\) with root \(r_w\).

\emph{Graph structure.}
For each operator \(n\), the signature:
\[
\sigma(n)=\textsf{canon}\!\big(\mathrm{name}(n),\mathrm{op}(n),\mathrm{target}(n),\mathrm{args}(n),\mathrm{kwargs}(n)\big),
\]
captures kind, attributes, and topology (edges are implied by arguments). The leaf hash is
\(
\mathcal{H}_g(n)=\Hsh\big(\sigma(n)\big)
\),
forming the structure tree \(\mathcal{T}_g\) with root \(r_g\). This construction enables logarithmic-depth proofs, making fast verification feasible even for large models.

To bind interfaces, we generate the commitment for inputs and outputs of \(S\):
\[
h_{D}=\Hsh\!\big(\big\|_{z\in D(S)}\Hsh(\textsf{canon}(z))\big), D\in\{\mathrm{In},\mathrm{Out}\}.
\]
Model configuration (percentile grid \(P\) and thresholds \(\tau\)) is part of the commitment and cannot change mid-dispute.

\Paragraph{Subgraph record and verification.} For each extracted \(S\), proposer publishes (i) its start/end \emph{indices} in the topological order, (ii) \(h_{\text{In}}\) and \(h_{\text{Out}}\), and (iii) Merkle inclusion proofs for all leaves referenced in \(S\) in \(\mathcal{T}_w\) and \(\mathcal{T}_g\). A challenger can verify by (1) recomputing \(r_w\) from revealed leaves using \(\mathcal{T}_w\) proofs, (2) checking membership of \(S\)'s operator signatures in \(\mathcal{T}_g\), and (3) recomputing \(h_{\text{In}},h_{\text{Out}}\) from observed tensors to check the integrity. Any tampering breaks a proof.

\subsection{Threshold-guided dispute localization}
\label{sec:iterative-subgraph-splitting}

In this section, we show how to deterministically select the offending model slice by comparing error-percentile profiles to calibrated per-operator thresholds.

\Paragraph{Why naive selection fails.}
Instruction/time-step bisection is undefined for GPUs; across inputs, intermediate tensor shapes may differ: without a shape-aware, deterministic selection rule, two honest parties may disagree about \emph{which} child is ``offending''. Thus, we adopt a tolerance-aware selection rule based on empirical error percentile thresholds. 

\Paragraph{Tolerance-aware selection rule.} Starting with $S_0\!=\!G$, proposer $P$ partitions $S_k$ into $N$ disjoint, contiguous slices $\{S_k^0,\ldots, S_k^{N-1}\}$. For each slice $S_k^j$ and live-out operator $v_i \in \mathrm{Out}(S_k^j)$, challenger $C$ re-executes $S_k^j$ and forms element-wise absolute and relative errors between proposer and its local outputs (flattened to 1-D):

\begingroup
\setlength{\abovedisplayskip}{-5pt}  
\begin{align*}
    \mathbf{e}_{\mathrm{abs}}^{(i)} = \text{flatten}(|y_i^{(P)} - y_i^{(C)}|), \;\mathbf{e}_{\mathrm{rel}}^{(i)} = \text{flatten}(\frac{|y_i^{(P)} - y_i^{(C)}|}{|y_i^{(C)}| + \epsilon}),
\end{align*}
\endgroup
and computes the percentile profile over
$p \in P = \{1,\allowbreak 5,\allowbreak 10,\allowbreak 15,\allowbreak \ldots,\allowbreak 95,\allowbreak 99,\allowbreak 100\}$:
\begingroup
\setlength{\abovedisplayskip}{0pt}  
\begin{align*}
    p_{\mathrm{abs/rel}}^{(i)}(p) &= \mathrm{Percentile}_p(\mathbf{e}_{\mathrm{abs/rel}}^{(i)}) \,.
\end{align*}
\endgroup

For each $v_i$, the challenger compares the observed error percentiles against the calibrated thresholds:
\begin{align}
    p_i^{(max)} &= \max_{p \in P} \left\{ \frac{p_{abs}^{(i)}(p)}{\tau_{abs}^{(i)}(p)}, \frac{p_{rel}^{(i)}(p)}{\tau_{rel}^{(i)}(p)} \right\}.
\end{align}
If any $p_i^{(max)} > 1$, the discrepancy lies within or upstream of \(S_k^j\) and the challenger continues the search.
$N$ way partition yields depth \(O(\log_{N} n)\).
The procedure stops when \(|S_k|=1\), isolating a single disagreeing operator \(v^*\) whose inputs and attributes are already fixed by prior commitments.

\subsection{Single-operator verification}
\label{sec:single-operator-verification}

\Paragraph{Routing policy.}
At the single operator $v^*$ (inputs and attributes already committed), the \emph{challenger} chooses one of two verification paths. \emph{Routing policy:} the challenger recomputes a reference $y_{v^*}^{\mathrm{ref}}$ and compares the elementwise error $|y_{v^*}^{P}-y_{v^*}^{\mathrm{ref}}|$ to the theoretical cap $\tau^{\text{theo}}_{v^*}$. If any element exceeds $\tau^{\text{theo}}_{v^*}$, it invokes path~(i); otherwise, it invokes path~(ii) to apply the tighter empirical thresholds.

\Paragraph{(i) Theoretical-bound verification.}
The challenger requests the proposer to submit a verifiable proof  that, given the committed inputs and operator metadata: (i) a canonical reference execution computes $y_{v^*}^{\mathrm{ref}}$ under IEEE-754 FP32 semantics; (ii) the operator-wise theoretical bound $\tau^{\text{theo}}_{v^*}$ is computed; (iii) element-wise, $\;|y_{v^*}^{P}-y_{v^*}^{\mathrm{ref}}|\le \tau^{\text{theo}}_{v^*}$. The coordinator verifies the single proof. If it passes the verification, the computation is \emph{accepted}; otherwise, the proposer \emph{loses} the dispute.

\Paragraph{(ii) Committee vote with empirical thresholds.}
Each member in the committee re-executes that operator to get $y_{v^*}^{\mathrm{ref}}$, computes element-wise absolute/relative errors versus $y_{v^*}^{P}$, forms the error percentile profile, and compare against the calibrated error percentile thresholds $\{\tau^{(*)}_{\mathrm{abs}},\tau^{(*)}_{\mathrm{rel}}\}$ for $v^*$. If the majority of committee members vote that the proposer's output is within the thresholds, it is \emph{accepted}; otherwise, the proposer \emph{loses} the dispute.

\vspace{-0.5em}
\subsection{Economic soundness and incentives}
\label{sec:economics}

Complementing the technical correctness of the dispute game, we design an economic mechanism on a blockchain coordinator to ensure incentive compatibility among rational proposers, challengers, and committee members. Adapting the fee-and-deposit pattern from optimistic accountability systems~\cite{kalodner2018arbitrum,zhang2024proof,sheng2024proof,zhao2025PoL}, \sys requires proposers and challengers to stake deposits $D_p$ and $D_{ch}$. A smart contract coordinator escrows these funds and automatically slashes (forfeiting the staked collateral) the losing party with $S_{\mathrm{slash}}$ upon adjudication. Meanwhile, committee members receive fixed compensation for participating in the arbitration process, regardless of the challenge outcome.

\Paragraph{Two detection channels: voluntary challenges vs.\ randomized audits.}
\sys supports two mutually exclusive supervisory channels per claim:
(i) \emph{a permissionless voluntary challenge} with probability $\phi_{ch}$, and
(ii) \emph{a randomized audit} triggered with probability $\phi$, where $\phi+\phi_{ch}\le 1$.
The costs for randomized audits are collected from users as service fees to ensure baseline security. In contrast, the cost of a voluntary challenge is borne by the challenger, where the reward for a successful challenge is set to strictly exceed this cost. Since voluntary challenges provide security redundancy without increasing the systematic fee burden, there is no rationale to exclude this cost-free oversight layer.

Let $\epsilon_1$ be the false negative rate (fraud missed due to tolerance allowances) and $\epsilon_2$ the false positive rate (wrong detection), where $\epsilon_1,\epsilon_2\in[0,1)$.
Then the probability that a cheating proposer is detected and penalized is
\begin{align}
d(\phi,\phi_{ch},\epsilon_1) \;=\; (\phi+\phi_{ch})\,(1-\epsilon_1).
\label{eq:detection-prob}
\end{align}
When $\epsilon_1$ is small, the detection probability is essentially the total supervision rate: $d(\phi,\phi_{ch},\epsilon_1)\approx \phi+\phi_{ch}$. For instance, with $\epsilon_1=0.001$, a combined challenge-and-audit rate of $5\%$ yields $d=4.995\%$, and $10\%$ yields $d=9.99\%$. As a representative operating point, if the coordinator targets $d^\star=0.05$, it suffices to ensure
\[
\phi+\phi_{ch}\ge \frac{0.05}{1-\epsilon_1}.
\]
Thus, under $\epsilon_1=0.001$, the required total supervision rate is only $0.05005$.

\begin{figure}[t]
    \centering
    \includegraphics[width=0.8\linewidth]{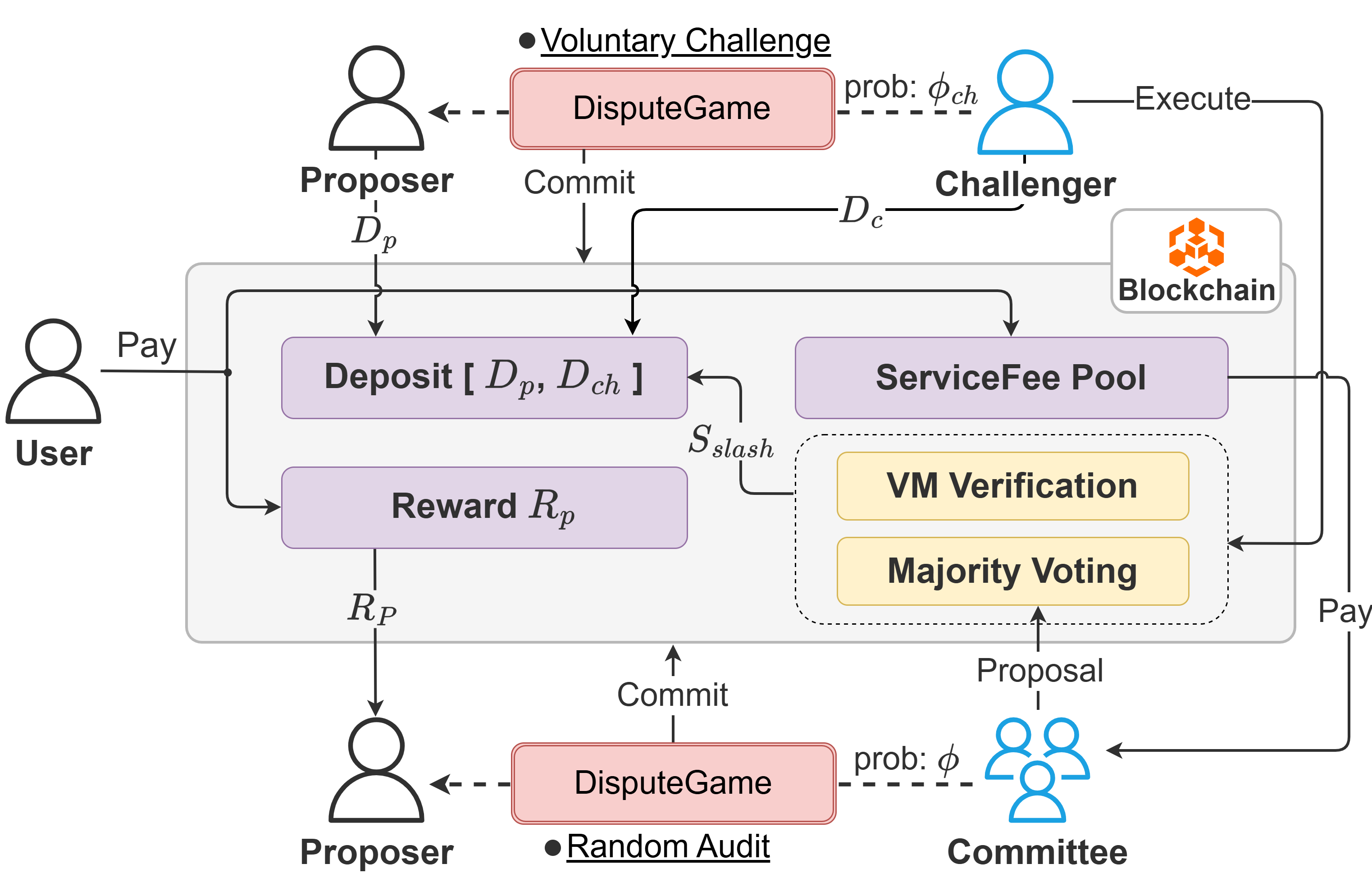}
    \vspace{-10pt}
    \caption{Mechanism overview (rewards, penalties, and two detection channels).}
    \vspace{-12pt}
    \label{fig:mechanism-overview}
    \Description{A diagram showing the mechanism overview with rewards, penalties, and two detection channels: voluntary challenges and randomized audits.}
\end{figure}

Fig.~\ref{fig:mechanism-overview} summarizes interactions in this mechanism.

\Paragraph{Incentive constraints.}
We choose penalties and deposits so that (a) honest execution dominates cheap cheating, (b) challenging is profitable when fraud exists but spamming is not, and (c) committee participation is sustainable.

\noindent\textbf{Proposer.}
The proposer faces three possible strategies: \textit{honest computation} ($h$), \textit{cheap cheating} ($c_1$, e.g., using a smaller model), and \textit{targeted cheating} ($c_2$, e.g., adversarial perturbation injection), with costs $C_p$, $C'_p$, and $C''_p$. Let $R_p$ be the task reward. Expected payoffs are
\begin{align}
u_p(h)   &= R_p - C_p - \epsilon_2 S_{\mathrm{slash}},\\
u_p(c_1) &= R_p - C'_p - d(\phi,\phi_{ch},\epsilon_1)\,S_{\mathrm{slash}},\\
u_p(c_2) &= R_p - C''_p.
\end{align}
Honest execution is individually rational when $R_p \ge C_p+\epsilon_2 S_{\mathrm{slash}}$. Moreover, to deter cheap cheating it suffices to make honesty strictly better than $c_1$:
\begin{align}
u_p(h) > u_p(c_1)
\quad\Longleftrightarrow\quad
S_{\mathrm{slash}} \;>\; \frac{C_p-C'_p}{d(\phi,\phi_{ch},\epsilon_1)-\epsilon_2},
\label{eq:S-lower}
\end{align}
assuming $d(\phi,\phi_{ch},\epsilon_1)>\epsilon_2$.
Note that for $c_2$, empirical evidence shows that effective targeted cheating requires $C''_p \gg R_p$, which implies $u_p(c_2)\le 0$.

\noindent\textbf{Voluntary challenger.}
Challengers freely select their verification strategy, which practically involves task re-execution and leaf verification at a total cost denoted by $C_{ch}$. A winning challenger receives a share $\alpha_{ch} \, S_{\mathrm{slash}}$ (where $\alpha_{ch} \in (0,1]$). The expected payoffs given the proposer is guilty/clean are
\begin{align}
u_{ch}\big|\text{guilty}
&= (1-\epsilon_1)\,\alpha_{ch} \, S_{\mathrm{slash}}
    - C_{ch} , \\
u_{ch}\big|\text{clean}
&=  - C_{ch}
    - (1-\epsilon_2)\, D_{ch}.
\end{align}
We have $u_{ch}|\text{clean}\le 0$ to deter spam. Conversely, a sufficient condition for honest challenges to be profitable is
\begin{align}
S_{\mathrm{slash}} > \frac{C_{ch}}{\alpha_{ch} \,(1-\epsilon_1)}.
\label{eq:S-for-challenger-worst}
\end{align}

\noindent\textbf{Audit committee.}
In a randomized audit, the coordinator samples $n$ committee members to adjudicate the claim. A sampled member $i$ incurs cost $C_a$. The coordinator pays $F_i$ (user service fee) to the committee member if the claim is ruled clean,
and pays a share $\alpha_{cm} \, S_{\mathrm{slash}}/n$ (where $\alpha_{cm} \in (0,1]$ and $\alpha_{cm} + \alpha_{ch} \le 1$) if guilt is found. Ex-post payoffs are
\begin{align}
u_{i}^{\text{cm}}\big|\text{ruled guilty} &= \tfrac{\alpha_{cm} \, S_{\mathrm{slash}}}{n}-C_a,\\
u_{i}^{\text{cm}}\big|\text{ruled clean}  &= F_i-C_a,
\end{align}
which are strictly positive whenever $ \frac{\alpha_{cm} \, S_{\mathrm{slash}}}{n} > C_a, \; F_i > C_a$.

\noindent\textbf{Nonempty feasible region for $S_{\mathrm{slash}}$.}
Let $\mathcal{L}\triangleq\max\{\mathcal{L}_1,\allowbreak \mathcal{L}_2,\allowbreak \mathcal{L}_3\}$, where
$\mathcal{L}_1=\frac{C_p-C'_p}{d(\phi,\phi_{ch},\epsilon_1)-\epsilon_2}$,
$\mathcal{L}_2=\frac{C_{ch}}{\alpha_{ch} \,(1-\epsilon_1)}$,
and $\mathcal{L}_3= \frac{n \, C_a}{\alpha_{cm}}$.
Assuming finite costs, and choosing knobs so that $d(\phi,\phi_{ch},\epsilon_1)>\epsilon_2$,
any $S_{\mathrm{slash}}\in(\mathcal{L}, D_p]$ satisfies all constraints and non-empty by setting $D_p>\mathcal{L}$.

\section{Implementation and evaluation}
\label{sec:implementation-and-evaluation}

We implement \sys as (i) a PyTorch runtime that instruments graphs, computes bounds, splits/executes subgraphs, advances dispute games, and emits/verifies commitments, and (ii) a set of smart contracts on Ethereum Holesky as a coordinator backend orchestrating commitments, dispute state, per-round bonds, and cost measurements. The runtime adds no custom kernels and runs unmodified models.

\subsection{Experimental setup}
\label{sec:eval-setup}

\Paragraph{Hardware and software configuration.}
Empirical error percentile threshold calibration uses four GPUs (NVIDIA RTX~4090, A100, H100, RTX~6000). We use PyTorch with FP32 forwards and FP64 for bound arithmetic. We set: fixed RNG seeds, fast-math disabled, CUDA/cuDNN library determinism, TF32/benchmark disabled, deterministic cuBLAS via \texttt{CUBLAS\_WORKSPACE\_CONFIG}. Contracts run on Ethereum Holesky testnet; we report on-chain computation costs in \emph{k} gas. We ignore network latency during interactions.

\Paragraph{Models and datasets.}
We evaluate \texttt{ResNet-152}~\cite{he2016deep} on ImageNet~\cite{deng2009imagenet}, \texttt{BERT-large}~\cite{devlin2019bert} on DBpedia~\cite{auer2007dbpedia}, \texttt{Qwen3-8B}~\cite{qwen3technicalreport} on C4~\cite{2019t5} (next-token prediction), and \texttt{Stable Diffusion v1-5}~\cite{Rombach_2022_CVPR}. PyTorch graph sizes range from $1k$ to $5k$ operators.

\Paragraph{Evaluation metrics.}
(i) \emph{Avg. dispute rounds}: average number of rounds required to reach the leaf; (ii) \emph{Avg. dispute time}: average off-chain dispute time (including both proposer and challenger with disk I/O); (iii) \emph{Avg. Merkle checks}: average number of weight/topology proof verifications; (iv) per-round \emph{substep time} during the dispute game (proposer partition vs.\ challenger selection). We inject perturbations separately into operators in the model and record per-round substep time to compute the variance; (v) \emph{DCR} (Dispute Compute Requirement): the challenger's total FLOPs to reach the leaf during the dispute game; and (vi) \emph{Cost Ratio}~$=~\mathrm{DCR}/$model forward FLOPs, for normalized comparison.

\subsection{System microbenchmarks and error analysis}

\begin{figure}
    \centering
    \includegraphics[width=\linewidth]{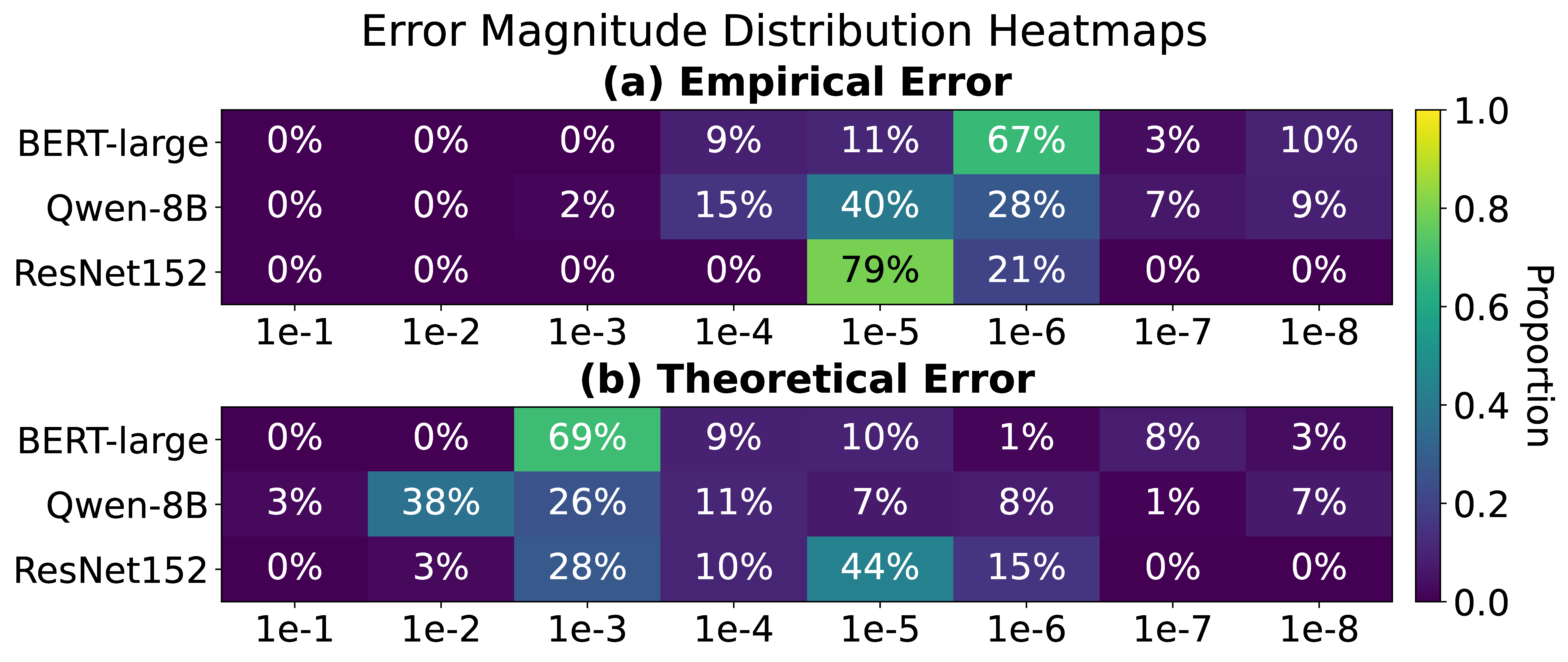}
    \vspace{-24pt}
    \caption{Error magnitude distribution heatmaps across ResNet 152, Qwen3-8B and BERT-large models.}
    \label{fig:empirical-theoretical-comparison}
    \Description{Heatmaps comparing empirical and theoretical error magnitude distributions for ResNet-152, Qwen3-8B, and BERT-large across all operators.}
    \vspace{-13pt}
\end{figure}

\begin{figure*}
    \centering
    \includegraphics[width=\linewidth]{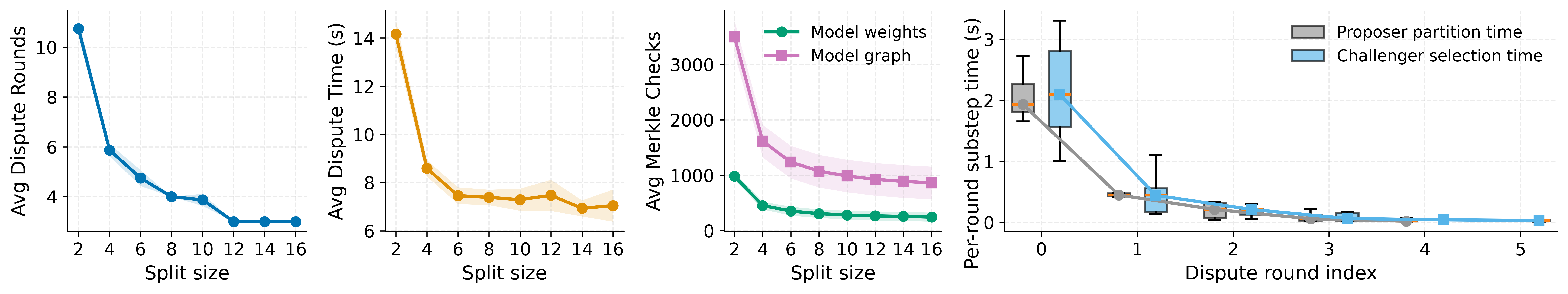}
    \vspace{-28pt}
    \caption{Microbenchmarks on \texttt{BERT-large} dispute game varying split size $N$. Left to right: avg dispute rounds (${\sim}O(\log_N|V|)$), avg dispute time, Merkle proof checks, and per-round substep latency.}
    \label{fig:system-microbenchmark}
    \Description{Four graphs showing BERT-large dispute game microbenchmarks: dispute rounds, dispute time, Merkle checks, and per-round substep latency vs.\ split size N.}
    \vspace{-5pt}
\end{figure*}

\Paragraph{Error-bound statistics and relative tightness.}
In Fig.~\ref{fig:empirical-theoretical-comparison} we collect element-wise empirical and theoretical errors across all operators for three models to compare distributions:

\Item{} \emph{BERT-large.} Mean empirical errors are extremely tight: $80\%$ of operators admit at most $10^{-6}$. The theoretical errors are much looser, with a mode at $10^{-3}$ covering $69\%$ of operators.

\Item{} \emph{Qwen3-8B.} The empirical errors concentrate in the $10^{-5}$–$10^{-6}$ bins, whereas theoretical errors are dominated by $10^{-2}$–$10^{-3}$, again yielding a $10^2$–$10^3$ gap.

\Item{} \emph{ResNet-152.} Both empirical and theoretical errors are around $10^{-5}$, but the latter shows a longer permissive tail.

\Paragraph{Implications for verification.}
Empirical checks are $10^2$--$10^3\times$ tighter than worst-case bounds for transformers, making evasion require consistently minute perturbations across many operators. Theoretical bounds alone leave more slack for LLMs, motivating the committee option at the leaf. Attack experiments confirm: empirical thresholds yield $0\%$ ASR while worst-case bounds allow a small window (Sec.~\ref{sec:attack-evaluation}).

\Paragraph{Dispute game statistics and scaling with $N$.}
Fig.~\ref{fig:system-microbenchmark} varies the partition size $N$. 
\emph{Avg. Dispute Rounds}: increasing $N$ reduces rounds from $\sim\!11$ at $N{=}2$ to $\sim\!3$ at $N{\ge}12$.
\emph{Avg. Dispute Time}: dispute time drops sharply from $N{=}2$ to $N{\approx}6$ and then flattens; the first round dominates latency because it handles the largest subgraph.
\emph{Merkle checks}: Merkle proof checks (graph and weights) shrink monotonically with $N$ (roughly $2{\times}$–$4{\times}$ reduction between $N{=}2$ and $N{\ge}8$).
\emph{Per-round substep latency}: the rightmost panel shows the time for the proposer's partition and the challenger's selection at dispute round index $i$ within an $N = 4$-way partition; both decay with $i$. Boxplot information is collected through eight dispute games across eight different perturbed operators through the model.
\emph{Guideline and insight}: $N\!\in\![8,12]$ minimizes wall-clock and coordinator verification work without creating oversized proofs.

\subsection{End-to-end cost and on-chain footprint}

\begin{table}[t]
\centering
\caption{Forward and dispute costs across models ($N = 2$). DCR is the challenger FLOP range to the leaf; \emph{Cost Ratio} = DCR/forward FLOPs. Gas sums all dispute interactions.}
\vspace{-5pt}
\label{tab:model-comparison}
\scriptsize
\setlength{\tabcolsep}{4pt}
\renewcommand{\arraystretch}{1.05}
\begin{tabular}{l *{4}{c}}
\toprule
\textbf{Metric} & \textbf{BERT} & \textbf{Diffusion} & \textbf{Qwen3-8B} & \textbf{ResNet} \\
\midrule
\rowcolor[HTML]{EFEFEF} 
Forward Cost (FLOPs, $10^{9}$) & 19.47 & 802.49 & 485.09 & 23.09 \\
Dispute Steps                            & 11    & 12     & 13     & 11    \\
\rowcolor[HTML]{EFEFEF} 
On-chain Cost (k gas)                    & 1984.4& 2073.1 & 2161.8 & 1984.4\\
DCR (FLOPs, $10^{9}$)     & [9.74, 20.6] & [316, 995] & [223, 485] & [11.7, 24.3] \\
\rowcolor[HTML]{EFEFEF} 
Cost Ratio Range                         & [0.50, 1.06]  & [0.39, 1.24]     & [0.46, 1.00]     & [0.51, 1.05] \\
\bottomrule
\end{tabular}
\vspace{-10pt}
\end{table}

\Paragraph{Compute and gas across models.}
Table~\ref{tab:model-comparison} summarizes compute and gas across four models in \sys.
In our $N=2$ setting, the 0.39–1.24$\times$ span arises because compute is unevenly distributed across operators in the canonical order: partitions covering heavier operators yield higher DCR, rather than from where the dispute’s disagreement happens to lie. There is no extra memory overhead beyond standard model/subgraph execution.
In the optimistic execution phase, the proposer posts a constant-size commitment (a state hash and a URI string) to the smart contract, which incurs a low, fixed on-chain cost of $\sim\!226$k gas. Additional gas is only consumed during dispute interactions, which scales with the number of verification rounds. For our evaluated models, dispute cost stays at $\sim\!2$M gas per game on Ethereum Holesky testnet.

\Paragraph{Deterministic execution overhead.}
The deterministic settings introduced in Sec.~\ref{sec:eval-setup} are also used during the optimistic execution phase (Phase~1 in Sec.~\ref{sec:system-architecture-and-workflow}) to ensure the system functions correctly. We test the overhead of these settings by running \texttt{Qwen3-8B} model on 100 WikiText-103~\cite{merity2017pointer} input samples. The experiments show that the software deterministic settings introduce a negligible overhead of ~0.3\% in terms of latency compared to the non-deterministic settings.

\Paragraph{Comparison to zkML approaches.}
Because most proof-based systems today arithmetize NN over finite fields and rely on integer/fixed-point encodings, direct head-to-head comparisons with our floating-point, native-kernel setting are inherently approximate: (i) \emph{Latency.} zkML-style systems have made rapid progress, yet still report per-inference proving times from tens of seconds (small CNNs/VGG/ResNet) to tens of minutes for LLM-scale models; verification is fast (seconds) with small proofs. By contrast, \sys incurs no proving phase: optimistic execution runs at native speed with negligible slowdowns from runtime flags (e.g., $0.3\%$ on \texttt{Qwen3-8B}), and disputes cost $\approx$0.39--1.24$\times$ a forward pass to reach and adjudicate the leaf (Table~\ref{tab:model-comparison}). (ii) \emph{Memory/footprint.} Proof systems may require very large prover RAM for LLM-scale circuits (e.g., up to $\lesssim$1~TB in zkML~\cite{chen2024zkml}); \sys adds \emph{no} memory beyond native subgraph execution and per-round Merkle proof checks. (iii) \emph{Precision.} Most zk pipelines quantize or otherwise encode floats into field-friendly integers; \sys verifies native IEEE-754 FP behavior by design. (iv) \emph{On-chain cost.} zk verifiers run off-chain in seconds with succinct proofs, whereas \sys incurs a minimal, constant on-chain fee ($\sim\!227$k gas) for optimistic commitments, and only pays for the interactive state machine and Merkle checks; for our models, disputes consume $\sim\!2$M gas with $N{=}2$.

\noindent\textbf{Where \sys fits.} When users accept tolerance-aware correctness, \sys offers orders-of-magnitude lower end-to-end latency and memory usage than zk proving for LLM-scale inference while preserving native kernels and heterogeneity. zk remains compelling when public verifiability with model secrecy is paramount; \sys is compelling when \emph{floating-point semantics and production performance} are constraints. 
\section{Discussion} \label{sec:discussion}

\Paragraph{Optimization compatibility.}
\label{sec:discussion-optimization-compatibility}
A practical advantage of \sys is that it is largely compatible with performance optimizations, as long as they preserve IEEE-754 semantics, including kernel-level optimizations such as tiling, unrolling, warp specialization, software pipelining, kernel fusion, instruction reordering, and parallel reduction reordering---all of which may change the rounding behavior but remain within the error bounds.
In contrast, \sys does \emph{not} aim to validate optimizations that change mathematical semantics or introduce input-dependent approximations (e.g., fast-math semantic changes, approximate math intrinsics via different hardware paths, undeclared quantization, early-exit strategies, and data-dependent pruning).
These optimizations can still be incorporated if they are made explicit in the commitment (i.e., treated as a distinct committed graph/semantic specification) and calibrated under the corresponding configuration class. Also, \sys is not fundamentally GPU-specific and can in principle extend to TPUs and other accelerators as long as the execution semantics are committed and re-executable for adjudication, with backend-specific calibration of tolerance thresholds for each hardware/software stack.

\Paragraph{Multi-step workloads and discrete decisions (tie-break rules).}
\sys naturally extends to multi-step settings (decoding, diffusion sampling, and training) by layering time over the dispute game: parties first bisect over committed step states to identify the earliest offending step, then localize within that step over the operator DAG to a leaf, yielding \emph{prefix finality}.
For discrete choices such as \texttt{argmax} or routing, where small tolerated logit differences may flip the outcome, \sys can enforce a \emph{deterministic, pre-committed tie-break rule}
(e.g., lexicographic tie-break; or a verifiable hash-seeded rule that deterministically selects among near-ties using a seed derived from committed public data), ensuring that honest executions converge on the same discrete decision while disputes remain localized and adjudicable.

\Paragraph{Comparison to replication.}
Replication is a natural baseline: always-on majority replication incurs a permanent $2\times$--$3\times$ compute multiplier, while our approach reduces this to $O(\phi C_p)$. Yet replication alone does not resolve adjudication under heterogeneous floating-point execution, where honest runs on different GPU stacks may diverge numerically because of IEEE-754 non-associativity and schedule-dependent reduction order. \sys instead commits to an operator-granular graph, localizes disagreement by threshold-guided partitioning, and reduces the terminal step to a single-operator check under either sound theoretical bounds or calibrated empirical thresholds. Its key advantage is therefore a principled and low-cost dispute path, offering significantly lower amortized overhead because arbitration only requires verifying a single operator rather than replaying the full model.


\Paragraph{Onboarding new configurations.}
A minor limitation is calibration churn: onboarding new devices, kernels, or library stacks may shift floating-point behavior outside previously committed empirical thresholds, potentially causing benign disputes.
Operationally, this can be mitigated by (i) whitelisting allowable optimization classes/configuration families for a deployment and calibrating under them, and (ii) treating newly observed benign drift as an onboarding event resolved by adjudication and followed by efficient threshold updates for the new configuration class.

\Paragraph{Limitations.}
\label{sec:limitations}
\sys is designed for settings where verification via re-execution is feasible, e.g., open-model deployments or permissioned/regulated environments in which authorized verifiers can reproduce execution for dispute resolution.
In practice, a regulated or contract-bound middle layer can enforce both service correctness and confidentiality.

\section{Related Work}
\label{sec:related-work}

\subsection{Verifiable Machine Learning}
Work on verifiable ML splits into cryptographic validity proofs and optimistic approaches. Zero-knowledge systems compile models to finite-field circuits and attach succinct proofs (e.g., SafetyNets~\cite{ghodsi2017safetynets}, zkML~\cite{chen2024zkml}, zkGPT~\cite{qu2025zkgpt}, zkllm~\cite{sun2024zkllm}), but still incurring substantial proving time/memory and often requiring quantization or restricted operators; TEE-assisted designs reduce proof cost but reintroduce trust in hardware vendors and exhibit performance degradation due to limited storage and compute capacity~\cite{tramerslalom, moon2025asgard,bai2025phantom,niu20223legrace}. Optimistic protocols (Agatha~\cite{zheng2021agatha}, opML~\cite{conway2024opml}) and refereed delegation (Verde~\cite{arun2025verde}) localize disputes on computation graphs and arbitrate on-chain, with practicality often hinging on replication or enforcing bitwise-reproducible operators (RepOps) across heterogeneous accelerators; recent systems generalize dispute games to Turing-complete programs and explore incentive mechanisms such as Proof-of-Sampling~\cite{mirkin2025arbigraph,zhang2024proof}. \sys takes a complementary tack: rather than enforcing bitwise equality, it verifies results up to principled, operator-level tolerances that acknowledge nondeterminism. Unlike approaches that determinize execution (e.g., high-precision rounding logs for verifiable training or RepOps libraries~\cite{srivastava2024optimistic,arun2025verde}), \sys aims to preserve native GPU kernels and heterogeneity while still providing portable accountability.

\vspace{-1em}
\subsection{Floating-Point Rounding Errors in Neural Networks}
A growing body of work shows that ignoring floating-point effects renders verification unsound in practice: real-number proofs can certify networks that fail under IEEE-754 execution, and deployed stacks exhibit nondeterminism from non-associative reductions, kernel fusion, and scheduler variation \cite{jia2021exploiting,szasz2025no}. Measurement tools such as FPRev reverse-engineer accumulation orders across software/hardware stacks~\cite{xie2025revealing}. In response, some systems try to \emph{eliminate} nondeterminism, e.g., reproducible operator libraries or training with high-precision states and logged rounding decisions to enable exact replay across GPUs, but at non-trivial performance and engineering cost \cite{arun2025verde,srivastava2024optimistic}.
\section{Conclusion} \label{sec:conclusion}
We introduce \sys, a tolerance-aware optimistic protocol that makes neural network computation verifiable without relying on bitwise determinism or trusted hardware.
\sys verifies results \emph{up to} operator-specific acceptance regions by combining portable IEEE-754 bounds with tight empirical error percentile thresholds, and it localizes disputes to a single operator via a Merkle-anchored dispute game followed by a low-cost leaf check.
This design preserves native kernels and performance while providing vendor-agnostic accountability; our implementation and attack study demonstrate the robustness and practicality of tolerance-aware verification for real-world ML workloads.

\balance
\newpage

\bibliographystyle{ACM-Reference-Format}
\bibliography{reference}

@article{ghodsi2017safetynets,
  title={Safetynets: Verifiable execution of deep neural networks on an untrusted cloud},
  author={Ghodsi, Zahra and Gu, Tianyu and Garg, Siddharth},
  journal={Advances in Neural Information Processing Systems},
  volume={30},
  year={2017}
}

@inproceedings{merity2017pointer,
  title={Pointer Sentinel Mixture Models},
  author={Merity, Stephen and Xiong, Caiming and Bradbury, James and Socher, Richard},
  booktitle={International Conference on Learning Representations},
  year={2017}
}

@inproceedings{auer2007dbpedia,
  title={Dbpedia: A nucleus for a web of open data},
  author={Auer, S{\"o}ren and Bizer, Christian and Kobilarov, Georgi and Lehmann, Jens and Cyganiak, Richard and Ives, Zachary},
  booktitle={international semantic web conference},
  pages={722--735},
  year={2007},
  organization={Springer}
}

@inproceedings{deng2009imagenet,
  title={Imagenet: A large-scale hierarchical image database},
  author={Deng, Jia and Dong, Wei and Socher, Richard and Li, Li-Jia and Li, Kai and Fei-Fei, Li},
  booktitle={2009 IEEE conference on computer vision and pattern recognition},
  pages={248--255},
  year={2009},
  organization={Ieee}
}

@inproceedings{zhang2019empirical,
  title={An empirical study of common challenges in developing deep learning applications},
  author={Zhang, Tianyi and Gao, Cuiyun and Ma, Lei and Lyu, Michael and Kim, Miryung},
  booktitle={2019 IEEE 30th international symposium on software reliability engineering (ISSRE)},
  pages={104--115},
  year={2019},
  organization={IEEE}
}

@inproceedings{shanmugavelu2024impacts,
  title={Impacts of floating-point non-associativity on reproducibility for HPC and deep learning applications},
  author={Shanmugavelu, Sanjif and Taillefumier, Mathieu and Culver, Christopher and Hernandez, Oscar and Coletti, Mark and Sedova, Ada},
  booktitle={SC24-W: Workshops of the International Conference for High Performance Computing, Networking, Storage and Analysis},
  pages={170--179},
  year={2024},
  organization={IEEE}
}

@inproceedings{chou2020deterministic,
  title={Deterministic atomic buffering},
  author={Chou, Yuan Hsi and Ng, Christopher and Cattell, Shaylin and Intan, Jeremy and Sinclair, Matthew D and Devietti, Joseph and Rogers, Timothy G and Aamodt, Tor M},
  booktitle={2020 53rd Annual IEEE/ACM International Symposium on Microarchitecture (MICRO)},
  pages={981--995},
  year={2020},
  organization={IEEE}
}

@article{choi2023tools,
  title={Tools for verifying neural models' training data},
  author={Choi, Dami and Shavit, Yonadav and Duvenaud, David K},
  journal={Advances in Neural Information Processing Systems},
  volume={36},
  pages={1154--1188},
  year={2023}
}

@inproceedings{jia2021proof,
  title={Proof-of-learning: Definitions and practice},
  author={Jia, Hengrui and Yaghini, Mohammad and Choquette-Choo, Christopher A and Dullerud, Natalie and Thudi, Anvith and Chandrasekaran, Varun and Papernot, Nicolas},
  booktitle={2021 IEEE Symposium on Security and Privacy (SP)},
  pages={1039--1056},
  year={2021},
  organization={IEEE}
}

@article{lahiany2022pteenet,
  title={PTEENet: Post-trained early-exit neural networks augmentation for inference cost optimization},
  author={Lahiany, Assaf and Aperstein, Yehudit},
  journal={IEEE Access},
  volume={10},
  pages={69680--69687},
  year={2022},
  publisher={IEEE}
}

@article{zhao2024atom,
  title={Atom: Low-bit quantization for efficient and accurate llm serving},
  author={Zhao, Yilong and Lin, Chien-Yu and Zhu, Kan and Ye, Zihao and Chen, Lequn and Zheng, Size and Ceze, Luis and Krishnamurthy, Arvind and Chen, Tianqi and Kasikci, Baris},
  journal={Proceedings of Machine Learning and Systems},
  volume={6},
  pages={196--209},
  year={2024}
}

@article{CanettiRivaRothblumIC13,
  author    = {Ran Canetti and Ben Riva and Guy N. Rothblum},
  title     = {Refereed Delegation of Computation},
  journal   = {Information and Computation},
  volume    = {226},
  pages     = {16--36},
  year      = {2013},
  doi       = {10.1016/j.ic.2013.03.003}
}

@inproceedings{CRR-CCS11,
  author    = {Ran Canetti and Ben Riva and Guy N. Rothblum},
  title     = {Practical Delegation of Computation Using Multiple Servers},
  booktitle = {Proceedings of the 18th ACM Conference on Computer and Communications Security (CCS)},
  pages     = {445--454},
  year      = {2011}
}

@inproceedings{GKR08,
  author    = {Shafi Goldwasser and Yael Tauman Kalai and Guy N. Rothblum},
  title     = {Delegating Computation: Interactive Proofs for Muggles},
  booktitle = {STOC},
  pages     = {113--122},
  year      = {2008}
}

@article{KRR22,
  author    = {Yael Tauman Kalai and Ran Raz and Ron D. Rothblum},
  title     = {How to Delegate Computations: The Power of No-Signaling Proofs},
  journal   = {Journal of the ACM},
  volume    = {69},
  number    = {1},
  articleno = {1},
  year      = {2022},
  doi       = {10.1145/3456867}
}

@inproceedings{esmaeili2024serene,
  title={SERENE: A collusion resilient replication-based verification framework},
  author={Esmaeili, Amir and Mtibaa, Abderrahmen},
  booktitle={2024 IEEE 13th International Conference on Cloud Networking (CloudNet)},
  pages={1--9},
  year={2024},
  organization={IEEE}
}

@inproceedings{zhao2005result,
  title={Result verification and trust-based scheduling in peer-to-peer grids},
  author={Zhao, Shanyu and Lo, Virginia and Dickey, C Gauthier},
  booktitle={Fifth IEEE International Conference on Peer-to-Peer Computing (P2P'05)},
  pages={31--38},
  year={2005},
  organization={IEEE}
}

@misc{pytorch-fx,
  author       = {{PyTorch Contributors}},
  title        = {{torch.fx}},
  howpublished = {\url{https://docs.pytorch.org/docs/main/fx.html}},
  note         = {PyTorch main documentation. Last updated: 2025-07-15. Accessed: 2025-09-23},
  year         = {2025},
  month        = jul
}

@misc{pytorch-reproducibility,
  author       = {{PyTorch Contributors}},
  title        = {Reproducibility},
  howpublished = {\url{https://docs.pytorch.org/docs/stable/notes/randomness.html}},
  note         = {PyTorch 2.8 documentation. Last updated: 2024-11-26. Accessed: 2025-09-23},
  year         = {2024},
  month        = nov
}

@misc{optimismcannonfpvm,
  author       = {{Optimism}},
  title        = {Multithreaded Cannon Fault Proof Virtual Machine},
  howpublished = {\url{https://specs.optimism.io/fault-proof/cannon-fault-proof-vm.html}},
  note         = {OP Stack Specification. Accessed: 2025-09-23},
  year         = {2025},
  month        = sep
}

@misc{optimismfaultproof,
  author       = {{Optimism}},
  title        = {Fault Proof},
  howpublished = {\url{https://specs.optimism.io/fault-proof/index.html}},
  note         = {OP Stack Specification. Accessed: 2025-09-23},
  year         = {2025},
  month        = sep
}

@article{kahan1996ieee,
  title={IEEE standard 754 for binary floating-point arithmetic},
  author={Kahan, William},
  journal={Lecture Notes on the Status of IEEE},
  volume={754},
  number={94720-1776},
  pages={11},
  year={1996}
}

@incollection{teutsch2024scalable,
  title={A scalable verification solution for blockchains},
  author={Teutsch, Jason and Reitwie{\ss}ner, Christian},
  booktitle={Aspects of Computation and Automata Theory with Applications},
  pages={377--424},
  year={2024},
  publisher={World Scientific},
}

@inproceedings{sun2024zkllm,
  title={zkllm: Zero knowledge proofs for large language models},
  author={Sun, Haochen and Li, Jason and Zhang, Hongyang},
  booktitle={Proceedings of the 2024 on ACM SIGSAC Conference on Computer and Communications Security},
  pages={4405--4419},
  year={2024}
}

@inproceedings{chen2024zkml,
  title={Zkml: An optimizing system for ml inference in zero-knowledge proofs},
  author={Chen, Bing-Jyue and Waiwitlikhit, Suppakit and Stoica, Ion and Kang, Daniel},
  booktitle={Proceedings of the Nineteenth European Conference on Computer Systems},
  pages={560--574},
  year={2024}
}

@inproceedings{qu2025zkgpt,
  title={zkGPT: An Efficient Non-interactive Zero-knowledge Proof Framework for LLM Inference},
  author={Qu, Wenjie and Sun, Yijun and Liu, Xuanming and Lu, Tao and Guo, Yanpei and Chen, Kai and Zhang, Jiaheng},
  booktitle={34th USENIX Security Symposium (USENIX Security 25)},
  year={2025}
}

@article{wang2023ezdps,
  title={ezDPS: An Efficient and Zero-Knowledge Machine Learning Inference Pipeline},
  author={Wang, Haodi and Hoang, Thang},
  journal={Proceedings on Privacy Enhancing Technologies},
  volume={2},
  pages={430--448},
  year={2023}
}

@inproceedings{tramerslalom,
  title     = {Slalom: Fast, Verifiable and Private Execution of Neural Networks in Trusted Hardware},
  author    = {Tramer, Florian and Boneh, Dan},
  booktitle = {International Conference on Learning Representations (ICLR)},
  year      = {2019},
  url       = {https://openreview.net/forum?id=rJVorjCcKQ}
}

@inproceedings{wei2018know,
  title={I know what you see: Power side-channel attack on convolutional neural network accelerators},
  author={Wei, Lingxiao and Luo, Bo and Li, Yu and Liu, Yannan and Xu, Qiang},
  booktitle={Proceedings of the 34th Annual Computer Security Applications Conference},
  pages={393--406},
  year={2018}
}

@article{niu20223legrace,
  title={3LegRace: Privacy-Preserving DNN Training over TEEs and GPUs},
  author={Niu, Yue and Ali, Ramy E and Avestimehr, Salman},
  journal={Proceedings on Privacy Enhancing Technologies},
  year={2022}
}

@inproceedings{bai2025phantom,
  title={Phantom: Privacy-Preserving Deep Neural Network Model Obfuscation in Heterogeneous TEE and GPU System},
  author={Bai, Juyang and Chowdhuryy, Md Hafizul Islam and Li, Jingtao and Yao, Fan and Chakrabarti, Chaitali and Fan, Deliang},
  year={2025},
  organization={34th USENIX Security Symposium}
}

@inproceedings{madry2018towards,
  title={Towards Deep Learning Models Resistant to Adversarial Attacks},
  author={Madry, Aleksander and Makelov, Aleksandar and Schmidt, Ludwig and Tsipras, Dimitris and Vladu, Adrian},
  booktitle={International Conference on Learning Representations},
  year={2018}
}

@inproceedings{moon2025asgard,
  title={ASGARD: Protecting On-Device Deep Neural Networks with Virtualization-Based Trusted Execution Environments},
  author={Moon, Myungsuk and Kim, Minhee and Jung, Joonkyo and Song, Dokyung},
  booktitle={Proceedings 2025 Network and Distributed System Security Symposium},
  year={2025}
}

@inproceedings{jia2021exploiting,
  title={Exploiting verified neural networks via floating point numerical error},
  author={Jia, Kai and Rinard, Martin},
  booktitle={International Static Analysis Symposium},
  pages={191--205},
  year={2021},
  organization={Springer}
}

@article{szasz2025no,
  title={No Soundness in the Real World: On the Challenges of the Verification of Deployed Neural Networks},
  author={Sz{\'a}sz, Attila and B{\'a}nhelyi, Bal{\'a}zs and Jelasity, M{\'a}rk},
  journal={arXiv preprint arXiv:2506.01054
        
        
        
        
        
        },
  year={2025}
}

@inproceedings{xie2025revealing,
  title={Revealing $\{$Floating-Point$\}$ Accumulation Orders in $\{$Software/Hardware$\}$ Implementations},
  author={Xie, Peichen and Gao, Yanjie and Wang, Yang and Xue, Jilong},
  booktitle={2025 USENIX Annual Technical Conference (USENIX ATC 25)},
  pages={1425--1440},
  year={2025}
}

@book{higham2002accuracy,
  title={Accuracy and stability of numerical algorithms},
  author={Higham, Nicholas J},
  year={2002},
  publisher={SIAM}
}

@inproceedings{kalodner2018arbitrum,
  title={Arbitrum: Scalable, private smart contracts},
  author={Kalodner, Harry and Goldfeder, Steven and Chen, Xiaoqi and Weinberg, S Matthew and Felten, Edward W},
  booktitle={27th USENIX Security Symposium (USENIX Security 18)},
  pages={1353--1370},
  year={2018}
}

@inproceedings{he2016deep,
  title={Deep residual learning for image recognition},
  author={He, Kaiming and Zhang, Xiangyu and Ren, Shaoqing and Sun, Jian},
  booktitle={Proceedings of the IEEE conference on computer vision and pattern recognition},
  pages={770--778},
  year={2016}
}

@inproceedings{devlin2019bert,
  title={Bert: Pre-training of deep bidirectional transformers for language understanding},
  author={Devlin, Jacob and Chang, Ming-Wei and Lee, Kenton and Toutanova, Kristina},
  booktitle={Proceedings of the 2019 conference of the North American chapter of the association for computational linguistics: human language technologies, volume 1 (long and short papers)},
  pages={4171--4186},
  year={2019}
}

@misc{qwen3technicalreport,
      title={Qwen3 Technical Report}, 
      author={Qwen Team},
      year={2025},
      eprint={2505.09388},
      archivePrefix={arXiv},
      primaryClass={cs.CL},
      url={https://arxiv.org/abs/2505.09388}, 
}

@InProceedings{Rombach_2022_CVPR,
    author    = {Rombach, Robin and Blattmann, Andreas and Lorenz, Dominik and Esser, Patrick and Ommer, Bj\"orn},
    title     = {High-Resolution Image Synthesis With Latent Diffusion Models},
    booktitle = {Proceedings of the IEEE/CVF Conference on Computer Vision and Pattern Recognition (CVPR)},
    month     = {June},
    year      = {2022},
    pages     = {10684-10695}
}

@article{goldberg1991every,
  title={What every computer scientist should know about floating-point arithmetic},
  author={Goldberg, David},
  journal={ACM computing surveys (CSUR)},
  volume={23},
  number={1},
  pages={5--48},
  year={1991},
  publisher={ACM New York, NY, USA}
}

@article{2019t5,
  author = {Colin Raffel and Noam Shazeer and Adam Roberts and Katherine Lee and Sharan Narang and Michael Matena and Yanqi Zhou and Wei Li and Peter J. Liu},
  title = {Exploring the Limits of Transfer Learning with a Unified Text-to-Text Transformer},
  journal = {arXiv e-prints},
  year = {2019},
  archivePrefix = {arXiv},
  eprint = {1910.10683},
}

@inproceedings{ansel2024pytorch,
  title={Pytorch 2: Faster machine learning through dynamic python bytecode transformation and graph compilation},
  author={Ansel, Jason and Yang, Edward and He, Horace and Gimelshein, Natalia and Jain, Animesh and Voznesensky, Michael and Bao, Bin and Bell, Peter and Berard, David and Burovski, Evgeni and others},
  booktitle={Proceedings of the 29th ACM International Conference on Architectural Support for Programming Languages and Operating Systems, Volume 2},
  pages={929--947},
  year={2024}
}

@inproceedings{merkle1987digital,
  title={A digital signature based on a conventional encryption function},
  author={Merkle, Ralph C},
  booktitle={Conference on the theory and application of cryptographic techniques},
  pages={369--378},
  year={1987},
  organization={Springer}
}

@manual{nvidia_cuda_programming_guide,
  title        = {CUDA C Programming Guide},
  author       = {{NVIDIA Corporation}},
  year         = {2024},
  note         = {Version 12.x},
  url          = {https://docs.nvidia.com/cuda/cuda-c-programming-guide/}
}

@article{zheng2021agatha,
  title={Agatha: Smart contract for DNN computation},
  author={Zheng, Zihan and Xie, Peichen and Zhang, Xian and Chen, Shuo and Chen, Yang and Guo, Xiaobing and Sun, Guangzhong and Sun, Guangyu and Zhou, Lidong},
  journal={arXiv preprint arXiv:2105.04919
        
        
        
        },
  year={2021}
}

@article{conway2024opml,
  title={opml: Optimistic machine learning on blockchain},
  author={Conway, KD and So, Cathie and Yu, Xiaohang and Wong, Kartin},
  journal={arXiv preprint arXiv:2401.17555
        
        
        
        },
  year={2024}
}

@article{zhang2024proof,
  title={Proof of Sampling: A Nash Equilibrium-Based Verification Protocol for Decentralized Systems},
  author={Zhang, Yue and Wang, Shouqiao and Tan, Sijun and Liu, Xiaoyuan and Moallemi, Ciamac C and Popa, Raluca Ada},
  journal={arXiv preprint arXiv:2405.00295
        
        
        
        },
  year={2024}
}

@article{srivastava2024optimistic,
  title={Optimistic verifiable training by controlling hardware nondeterminism},
  author={Srivastava, Megha and Arora, Simran and Boneh, Dan},
  journal={Advances in Neural Information Processing Systems},
  volume={37},
  pages={95639--95661},
  year={2024}
}

@article{arun2025verde,
  title={Verde: Verification via refereed delegation for machine learning programs},
  author={Arun, Arasu and Arnaud, Adam St and Titov, Alexey and Wilcox, Brian and Kolobaric, Viktor and Brinkmann, Marc and Ersoy, Oguzhan and Fielding, Ben and Bonneau, Joseph},
  journal={arXiv preprint arXiv:2502.19405
        
        
        
        },
  year={2025}
}

@InProceedings{sheng2024proof,
  author    = {Sheng, Peiyao and Rana, Ranvir and Bala, Senthil and Tyagi, Himanshu and Viswanath, Pramod},
  title     = {{Proof of Diligence: Cryptoeconomic Security for Rollups}},
  booktitle = {6th Conference on Advances in Financial Technologies (AFT 2024)},
  pages     = {5:1--5:24},
  series    = {Leibniz International Proceedings in Informatics (LIPIcs)},
  volume    = {316},
  year      = {2024},
  publisher = {Schloss Dagstuhl -- Leibniz-Zentrum f{\"u}r Informatik},
  doi       = {10.4230/LIPIcs.AFT.2024.5},
}

@article{mirkin2025arbigraph,
  title={Arbigraph: Verifiable Turing-Complete Execution Delegation},
  author={Mirkin, Michael and Chen, Hongyin and Eitan, Ohad and Granot, Gal and Eyal, Ittay},
  journal={Cryptology ePrint Archive},
  year={2025}
}

@article{higham2019new,
  title={A new approach to probabilistic rounding error analysis},
  author={Higham, Nicholas J and Mary, Theo},
  journal={SIAM journal on scientific computing},
  volume={41},
  number={5},
  pages={A2815--A2835},
  year={2019},
  publisher={SIAM}
}

@misc{zhao2025PoL,
      title={Proof-of-Learning with Incentive Security}, 
      author={Zishuo Zhao and Zhixuan Fang and Xuechao Wang and Xi Chen and Hongxu Su and Haibo Xiao and Yuan Zhou},
      year={2025},
      eprint={2404.09005},
      archivePrefix={arXiv},
      primaryClass={cs.CR},
      url={https://arxiv.org/abs/2404.09005}, 
}

\newpage

\appendix
\section{Floating-Point}

\subsection{Floating-Point Basics}
\label{sec:fp-basic}

We assume IEEE-754 arithmetic with rounding to nearest, ties to even, and exclude overflow/underflow/NaNs unless stated otherwise.
Let $u$ denote the \emph{unit roundoff} (machine epsilon divided by~2), e.g.\ $u=2^{-24}$ for \texttt{float32}.
For any basic operation $\circ\in\{+,-,\times,/\}$ on normalized operands, the \emph{standard model}
captures one rounding step as
\begin{equation}
\label{eq:standard-model}
\operatorname{fl}(x \circ y) \;=\; (x \circ y)(1+\delta), \qquad |\delta|\le u.
\end{equation}
It is often convenient to write products of many $(1+\delta)$ terms as
\begin{equation}
\prod_{j=1}^{m}(1+\delta_j) \;=\; 1 + \theta_m,
\qquad |\theta_m|\le \gamma_m
\quad\text{with}\quad
\gamma_m \;\triangleq\; \frac{m u}{1-m u},
\label{eq:gamma-def}
\end{equation}
valid for $m u<1$; see~\cite[Sec.\,2--3]{higham2002accuracy,goldberg1991every}.  We will also use that $\gamma_k$ is monotone in $k$ and that
$\gamma_k = k u + O(u^2)$ for fixed $k$ and small~$u$.

\Paragraph{Modeling conventions.}
Unless stated otherwise, our bounds assume:
(i) rounding to nearest, ties to even; (ii) no subnormal underflow/overflow; (iii) each rounding error $\delta$ obeys $|\delta|\le u$ as in~\eqref{eq:standard-model}.
When we invoke a probabilistic model, we additionally assume independent, mean-zero, bounded rounding errors as stated below.

\subsection{Theoretical error bounds}

In this section, we elaborate on the derivation of theoretical error bound formula for floating-point operations, including the deterministic and probabilistic bounds.

Let $u$ be the unit roundoff (e.g., $u=2^{-24}$ for \texttt{float32}). In the standard model~\cite{higham2002accuracy,goldberg1991every},
\[
\operatorname{fl}(x\circ y) \;=\; (x\circ y)(1+\delta), 
\qquad |\delta|\le u,\quad \circ \in \{+,-,\times,/\},
\]

For an $n$-term sum $s=\sum_{i=1}^n x_i$ evaluated by sequential FP adds,
\[
\widehat{s}_k=\operatorname{fl}(\widehat{s}_{k-1}+x_k)=(\widehat{s}_{k-1}+x_k)(1+\delta_k),\quad k=1,\dots,n,\quad \widehat{s}_0=0,
\]
one can unroll to obtain
\begin{align}
\label{eq:sum-unroll}
\widehat{s}
=\sum_{i=1}^n x_i \prod_{j=i+1}^{n} (1+\delta_j).
\end{align}
Taking absolute values and using the triangle inequality,
\begin{align}
|\widehat{s}-s|
\;\le\;
\sum_{i=1}^n |x_i|\;\Big|\prod_{j=i+1}^n (1+\delta_j)-1\Big|.
\end{align}

We use the tail-product bound~\cite[Sec.\,3.1]{higham2002accuracy}: if $|\delta_j|\le u$ and $m u<1$, then
\begin{align}
\Big|\prod_{j=1}^{m}(1+\delta_j)-1\Big|
\;\le\;
\gamma_m \;\triangleq\; \frac{m u}{1-m u}.
\end{align}
Applying the lemma to each tail with $m=n-i$ and using that $\gamma_m$ increases with $m$ yields the deterministic bound
\begin{equation}
\label{eq:det-sum}
|\widehat{s}-s|
\;\le\;
\gamma_{n-1}\,\sum_{i=1}^n |x_i|,
\qquad \text{valid when }(n-1)u<1.
\end{equation}

Following probabilistic model~\cite{madry2018towards}, assume the rounding errors in a length-$n$ computation are independent, mean-zero, and bounded by $u$. Then for any $\lambda>0$,
\begin{align}
\prod_{i=1}^{n} (1+\delta_i) \;&=\; 1+\theta_n,\\
|\theta_n| \;&\le\; \widetilde{\gamma}_n(\lambda) 
\;\triangleq\; \exp\!\Big(\lambda \sqrt{n}\,u \;+\; \tfrac{n u^2}{1-u}\Big) - 1,
\end{align}
with probability at least
\begin{align}
P(\lambda)\;=\;1 - 2\exp\!\Big(-\tfrac{\lambda^2(1-u)^2}{2}\Big).
\end{align}

Since $e^t \le \tfrac{1}{1-t}$ for $0\le t<1$, we have

\begin{align}
\widetilde{\gamma}_n(\lambda) \;\le\; \lambda \sqrt{n}\,u + O(u^2), \qquad \lambda \sqrt{n}\,u + \tfrac{n u^2}{1-u} < 1.
\end{align}

Applied to summation~\eqref{eq:det-sum}, we obtain the high-probability bound
\begin{align}
\Pr\!\left(|\widehat{s}-s|\;\le\;\widetilde{\gamma}_{n-1}(\lambda)\sum_{i=1}^n |x_i| \right)\;\ge\;P(\lambda).
\end{align}

\subsection{Implemented PyTorch operation list for theoretical error bounds}

We elaborate all the PyTorch operations for which we derive theoretical error bounds in both deterministic and probabilistic models. Operators include: basic arithmetic and elementwise functions (Add, Sub, Mul, Div, Pow, Neg, Sqrt, Rsqrt, Exp, Log, Sin, Cos, Tanh); activations (ReLU, GELU, SiLU); normalization and softmax (Softmax, BatchNorm, LayerNorm, GroupNorm); linear algebra and convolutions (Matmul, Bmm, Linear, Conv2D); reductions, pooling, upsampling (Mean, AdaptiveAvgPool2d, MaxPool2d, Interpolate); structural and non-arithmetic ops (Concatenate, Slice, Flatten, Indexing, Views/transforms, Type/device/movement, Masked fill, Embedding, Max/Min, Dropout eval).

\section{Stability of Empirical Error Percentile Profiles}
\label{sec:statistical-validation}

\Paragraph{Goal.}
We assess the stability of per\mbox{-}operator empirical error percentile profiles as additional samples accrue. Let
$\mathcal{I}$ index operators, $\mathcal{P}$ the set of error percentiles (e.g., $p\in\{10,15,\dots,90\}$), and $n$ the sample size (e.g., $n=50$).
For operator $i\in\mathcal{I}$ and percentile $p\in\mathcal{P}$, denote the per\mbox{-}sample sequence by $\{y_{i,p,t}\}_{t=1}^n$. The running median is
\begin{equation}
\widetilde{\theta}_{i,p}(k) \;=\; \operatorname{median}\big\{y_{i,p,1},\dots,y_{i,p,k}\big\},\qquad k=1,\dots,n,
\end{equation}
with $\theta_{i,p}:= \widetilde{\theta}_{i,p}(n)$ the point estimate at $n$.
Unless stated otherwise, the median of an even number of points is the midpoint of the two central order statistics.

\Paragraph{Relative scale.}
To obtain scale\mbox{-}free magnitudes we use the symmetric relative change
\begin{equation}
\delta(a,b) \;=\; \frac{2\,|a-b|}{|a|+|b|+\varepsilon},
\qquad \varepsilon>0 \ \text{small}.
\label{eq:symmetric-rel-change}
\end{equation}
When normalizing a quantity $x$ by a baseline $z$ we use $|x|/(|z|+\varepsilon)$ with the same safeguard $\varepsilon$.

\subsection{Diagnostics (per operator and per percentile)}
We compute four complementary robustness diagnostics for each $(i,p)$, reported in relative units. Throughout, assume $W\in\{1,\dots,n-1\}$ and exclude non\mbox{-}finite values before computing any statistics.

\begin{enumerate}[label=(D\arabic*), leftmargin=1.5em]
\item \textbf{Short\mbox{-}horizon relative drift (\emph{SupNorm}).}
Detects drift of the running median within the last $W$ steps:
\begin{equation}
\mathrm{SupNorm}_{i,p}
\;=\;
\max_{k\in\{n-W,\dots,n-1\}}
\delta\!\big(\widetilde{\theta}_{i,p}(n),\,\widetilde{\theta}_{i,p}(k)\big).
\end{equation}
(This matches the usual supremum interpretation over the most recent horizon. If a single\mbox{-}lag version is preferred, replace the max by $k=n-W$.)

\item \textbf{Maximum leave\mbox{-}one\mbox{-}out influence (\emph{Jackknife}).}
Measures sensitivity to any single sample:
\begin{equation}
\mathrm{JK}_{i,p}
\;=\;
\max_{1\le t\le n}\;
\frac{\big|\operatorname{median}(\{y_{i,p,s}\}_{s\neq t})-\theta_{i,p}\big|}
{|\theta_{i,p}|+\varepsilon}.
\end{equation}
(Optionally replace the numerator with $\delta\!\big(\operatorname{median}(\{y_{i,p,s}\}_{s\neq t}),\allowbreak\,\theta_{i,p}\big)$ for full symmetry.)

\item \textbf{Tail adjustment over the last $W$ steps (\emph{TailAdj}).}
Detects lingering adjustments in the tail of the running\mbox{-}median curve:
\begin{equation}
\mathrm{TailAdj}_{i,p}
\;=\;
\max_{k\in\{n-W,\dots,n-1\}}
\frac{\big|\widetilde{\theta}_{i,p}(k\!+\!1)-\widetilde{\theta}_{i,p}(k)\big|}
{|\theta_{i,p}|+\varepsilon}.
\end{equation}

\item \textbf{Rolling\mbox{-}window variability (\emph{RollSD}).}
Captures local variability of windowed medians. Let $r_{i,p}(k)$ be the median of a length\mbox{-}$W$ sliding window ending at $k$ (defined for $k=W,\dots,n$). Report
\begin{equation}
\mathrm{RollSD}_{i,p}
\;=\;
\frac{\operatorname{sd}\big(\{r_{i,p}(k)\}_{k=W}^{n}\big)}
{|\theta_{i,p}|+\varepsilon}.
\end{equation}
\end{enumerate}

\subsection{Global drift across percentiles (per operator)}
To summarize cross\mbox{-}percentile instability, each operator is represented by the worst\mbox{-}case short\mbox{-}horizon drift over the last $W$ steps:
\begin{equation}
\Delta_i^{(\infty)}
\;=\;
\max_{p\in\mathcal{P}}
\ \max_{k\in\{n-W,\dots,n-1\}}
\delta\!\big(\widetilde{\theta}_{i,p}(n),\,\widetilde{\theta}_{i,p}(k)\big).
\label{eq:supnorm-across-p}
\end{equation}

\subsection{Aggregation and reporting}
All diagnostics are computed per operator and per percentile and then aggregated across operators. For each percentile $p$ (i.e., each column \texttt{abs\_q$p$}), we report the median (50th) and upper\mbox{-}decile (90th) across operators.

\Paragraph{Default settings.}
Unless stated otherwise, we use $W=10$ (tail/window length), $n=50$ samples per operator, and $\mathcal{P}=\{10,15,\dots,90\}$. Non\mbox{-}finite values are excluded prior to any per\mbox{-}operator summaries. All normalizations use \eqref{eq:symmetric-rel-change} with the same $\varepsilon$.

\end{document}